\documentclass[a4paper,12pt]{article}

\usepackage{amsmath}
\usepackage{graphicx}
\usepackage{hyperref}
\usepackage{color}
\usepackage{subfigure}
\usepackage{cite}
\usepackage{slashed}
\usepackage{textcomp}
\usepackage{amssymb}

\numberwithin{equation}{section}

\def \mphi{M_{\phi}}
\def \yS{y_S}
\def \yP{y_P}
\def \mZ{M_{Z'}}

\def \gX{g_X}
\def \mA{M_A}
\def \gA{\Gamma_A}
\def \gVq{g_V^q}
\def \gAq{g_A^q}
\def \gVt{g_V^t}
\def \gAt{g_A^t}

\def\ltap{\, \raisebox{-.4ex}{\rlap{$\sim$}} \raisebox{.4ex}{$<$} \,} 
\def\gtap{\, \raisebox{-.4ex}{\rlap{$\sim$}} \raisebox{.4ex}{$>$} \,} 
\def\beq{\begin{equation}} 
\def\eeq{\end{equation}} 
\def\barr{\begin{array}}
\def\earr{\end{array}}
\def\dis{\displaystyle}
\def\gev{\, {\rm GeV}} 
\def\tev{\, {\rm TeV}} 
\def\sq{{\cal S_Q}}
\def\stp{{\cal S_T^+}}
\def\stm{{\cal S_T^-}}
\def\pq{{\cal P_Q}}
\def\pt{{\cal P_T}}
\def\pv{{\cal P_V}}
\def\pa{{\cal P_A}}
\def\lt{\lambda_t}
\def\lbt{\lambda_{\bar t}}

\begin{document}

\begin{center} 

\begin{flushright}
{CERN-PH-TH/2010-314}\\
\end{flushright}

\vspace*{40pt}

{\large\bf Top polarization, forward-backward asymmetry and new physics}\\[3ex]

Debajyoti Choudhury$^a$, Rohini M. Godbole$^{b,}$\footnote{Permanent Address:
Centre for High Energy Physics, Indian Institute of Science, Bangalore 560 012, India}, Saurabh D. Rindani$^c$, Pratishruti Saha$^a$\\

\vspace*{5pt}

\begin{footnotesize}
$^a$ {\sl Department of Physics and Astrophysics, University of Delhi, 
Delhi 110 007, India.}\\[1ex]
$^b$ {\sl Theory Unit, CERN, CH-1211, Geneva 23, Switzerland.}\\[1ex]
$^c$ {\sl Theoretical Physics Division, Physical Research Laboratory, 
Navrangpura, Ahmedabad 380 009, India.}\\[1ex]
\end{footnotesize}

\end{center} 

\vspace*{40pt}
 
\begin{abstract} 
\noindent
We consider how the measurement of top polarization at the Tevatron
can be used to characterise and discriminate among different new physics
models that have been suggested to explain the anomalous top forward-backward
asymmetry reported at the Tevatron. This has the advantage of catching the
essence of the parity violating effect characteristic to the different 
suggested new physics models. Other observables constructed from these 
asymmetries are shown to be useful in discriminating between the models, 
even after taking into account the statistical errors. Finally, we 
discuss some signals at the 7 TeV LHC.

\vspace*{40pt}
\noindent
\texttt{PACS Nos:14.65.Ha,13.88.+e} \\ 
\texttt{Key Words:top,asymmetry,polarization}
\end{abstract}

\vspace*{40pt}

\newpage
\section{Introduction}
\label{sec:intro}

The study of the third generation quarks in the Standard Model(SM)
fermions continues to throw up surprises. 
Be it a $\sim 3\sigma$ deviation in $A^b_{FB}$ in the SM fit to 
the electroweak(EW) precision measurements at LEP~\cite{Nakamura:2010zzi}, 
a forward-backward asymmetry in top-pair production~\cite{CDF_AFB_2008_PRL,
D0_AFB_2008_PRL,CDF_AFB_2009,CDF_AFB_2010,D0_AFB_2010}
significantly larger than what the SM predicts 
or the discrepancy in the value of the $B_s^0$--$\overline{B_s^0}$ 
mixing as suggested by the recent measurement of the inclusive 
dimuon asymmetry~\cite{Abazov:2010hv} at the Tevatron, some
deviations from the SM remain.
While, individually, none of them are large enough to merit
the status of a discovery of physics beyond the SM, they are,
nonetheless, intriguingly poised to warrant serious attention. 
This is especially so on account of the largeness of the
third generation fermion masses. With the top quark mass scale being
very close to the EW scale, it is quite conceivable that the third
generation (or, at least, the top quark) plays an important role in
electroweak symmetry breaking itself. Indeed, very many ideas going
beyond the SM quite often also predict discernible new physics effects
for processes involving the third generation. On the experimental front, 
constraints on the universality of interactions are far
less restrictive for these fermions as opposed to those for the first two
generations. Hence, on very generic grounds, top quark physics studies
at the Tevatron and the LHC, hold potential for probing 
new physics~\cite{top_reviews}. It is, 
therefore, a small wonder that very many ``explanations'' of the physics
responsible for the aforementioned discrepancies have been offered and
a majority of them accord a special role to the third generation. 
In this article, we concentrate on the anomaly in the forward-backward
(FB) asymmetry in top-pair production.

Within the SM, the dominant $t \bar t$ production mode at the Tevatron
is a pure QCD one and is FB symmetric at the tree-level.
Weak interactions are highly subdominant and give a very small 
asymmetry ($A_{FB} \ll 1\%$).
It is the interference of the tree-level QCD amplitudes and
higher order terms that yields the dominant SM contribution to this
asymmetry and results in $A_{FB} \sim 5\%$~\cite{AFB_SMNLO,AFB_SM_others}. 
This has led to several authors~\cite{Sehgal:1987wi,DC&RMG} 
advocating the measurement of $A_{FB}$ as a tool for probing physics 
beyond the SM. 

The CDF and the D0 experiments at the Tevatron kindled a great deal of
interest in top-pair production by reporting a FB asymmetry ($A_{FB}$)
of 17\%~\cite{CDF_AFB_2008_PRL} and 12\%~\cite{D0_AFB_2008_PRL}
respectively\footnote{The measurements reported by D0 are uncorrected for 
kinematic acceptance.}. 
Many possible New Physics (NP) scenarios have been offered as explanations. 
In a later update, CDF revised this number to 19.3\%.~\cite{CDF_AFB_2009}  
Although the analysis of more data has reduced the significance 
of the CDF result to about $1.8\sigma$ ($A_{FB}$ = 15\%)~\cite{CDF_AFB_2010}
and D0 has presented a newer value of 8\%~\cite{D0_AFB_2010}, the deviation from the SM is
still non-negligible and, coupled with the other longstanding
discrepancies in certain third-generation observables, continues to
elicit much interest~\cite{Murayama,Tait,Frampton,AFB_others,Chivukula:2010fk,Jung:2009pi}.

Forward-backward asymmetry, at the Tevatron, is defined as
\beq
A_{FB} = \dfrac{\sigma(\cos\theta_t > 0) - \sigma(\cos\theta_t < 0)}
                {\sigma(\cos\theta_t > 0) + \sigma(\cos\theta_t < 0)}
\eeq
where $\theta_t$ is the angle made by the top quark with the
direction of the proton in the lab-frame.  Experimentally, the
measurement is made in the semi-leptonic channel, where the angle
($\theta_h$) made by the hadronically decaying top(anti-top) with the
proton beam and the charge $Q_l$ of the decay lepton from the
anti-top(top) are together used to construct the net top current in
the direction of the proton beam. $\cos\theta_t$ above is thus
equivalent to $-Q_l\cdot\cos\theta_h$~\cite{CDF_AFB_2009}. 
CP invariance is assumed to hold good.
 
Naively, a non-zero $A_{FB}$ seems to be an indication of some
violation of a discrete symmetry and, indeed, most models that purport
to explain this anomaly have invoked a parity-violating interaction
for the top-quark. While this assumption certainly holds true for 
any $s$-channel NP contribution to $q \bar q \to t \bar t$, 
clearly, it is not applicable when $t$- or $u$-channel contributions
are present as well. In other words, the measured $A_{FB}$, in the 
presence of any NP interactions, may accrue from either explicit 
parity violation(dynamics) or the effects of $t$-($u$)-channel
propagators(kinematics) or a combination of both.
We believe this issue has not been stressed sufficiently in the literature.
The new physics explanations have differing levels of parity violation
encoded in the chiral structure of the interactions. As this is reflected
in the $A_{FB}$ to varying amounts, it would be interesting to construct
a probe thereof. The polarization of a single top offers one such probe. 
In fact, this observation was first made in Ref.~\cite{Hikasa:1999wy} in 
the context of sfermion exchange contributions to $t \bar t$ production 
in $R$-parity violating MSSM (an analog of this model is one of the 
candidates for an explantaion of the observed $A_{FB}$). 

It is a thankful coincidence that the top quark polarization 
is also a quantity that is amenable to an experimental measurement
due to the large mass of the $t$. Being heavy, the top quark decays
before it hadronizes and thus the decay products carry the memory
of its spin direction. This correlation between the kinematic 
distribution of the decay products and the top spin direction, 
can be used to get information about the latter.  
In fact, many studies exploring the use of the polarization of the 
top quark as a probe and discriminator of new physics~\cite{toppol_bsm},
as a means to sharpen up the signal of new physics~\cite{Agashe:2006hk} and
to obtain information on  $t \bar t$ production 
mechanism~\cite{Hikasa:1999wy,jhepus2,Godbole:2010kr} exist in literature. 
Different probes of the top polarisation which use the above 
mentioned correlation have been constructed~\cite{Godbole:2010kr,Godbole:2006tq,Shelton},
the angular distribution of the decay leptons providing a particularly 
robust probe due to its insensitivity to higher order corrections~\cite{LEP_QCD}
and to possible new physics in the $tbW$ vertex~\cite{Hioki_et_al,Rindani,jhepus1}.

A single-top polarization asymmetry($A_P$) can be defined as
\beq
A_P = \dfrac{\sigma(+) - \sigma(-)}{\sigma(+) + \sigma(-)} \ ,
\eeq
where $+$ or $-$ denote the helicity of the top quark and the
helicities of the $\bar t$ are summed over. The SM prediction for this
arises due to EW effects and is expected to be small.
In this note, we explore the predictions for $A_P$ originating from
the different new physics explanations of the FB asymmetry.  
Spin polarization studies are a part of the agenda at the Tevatron
as well as the LHC~\cite{LHC_spin_plans}. 
Both the CDF~\cite{CDF_spincorr} and the D0~\cite{D0_spincorr} 
experiments at the Tevatron have reported measurements of 
spin correlation coefficients.  
In addition, CDF also reports a measurement of $t \bar t$ 
helicity fractions~\cite{CDF_spincorr}. We would like to point out
that, while these observables may also be able to distinguish between
the different NP scenarios under consideration, they involve
measurement of the polarization of the top as well as the anti-top and
hence, are experimentally more challenging. This is underscored by the
fact that the aforementioned measurements are accompanied by fairly
large error bars.  On the other hand, measurement of $A_P$ requires
knowledge of the polarization of only the top and thus provides an
advantage in terms of the statistics that may be obtained.

The rest of the article is organised as follows. In the next section, we 
discuss the rudiments of the models we use as templates, with particular 
emphasis on the features that are germane to the issue at hand. This 
is followed, in section \ref{sec:analytic}, by a comparative discussion
of the structure of some observables and their efficacy in distinguishing
certain features. The numerical results, pertaining to the resolving powers
of various observables, are presented in section \ref{sec:numeric}. Finally, we 
summarise in section \ref{sec:summary}.

\section{Model Templates}
\label{sec:models}

Rather than probe each and every model that has been proposed, we select
some that, to our mind, serve as templates. 
Broadly speaking, four classes of explanations have been suggested.
The first two involve new $t$-channel exchanges in $q \bar q \to t\bar t$
while the third involves s-channel contributions to the same. 
For very large masses of the exchanged bosons, all three reduce to 
four-fermion contact interactions which constitute the fourth category.
We do not discuss the last-mentioned explicitly as it can be approached
from each of the other cases in the appropriate limit\footnote{In principle,
there could be scenarios wherein more than one such NP effect could play a role, 
albeit to different degrees. Once again we desist from discussing these
explicitly as the gross features thereof can be deduced by judiciously 
combining the templates that we do examine hereafter.}.
The particle exchanged in the $t$-channel could either be a scalar or a 
vector and we shall discuss an example of each. 
Scalar exchanges in the $s$-channel do not contribute
to $A_{FB}$. On the other hand, $s$-channel vector exchanges are fairly
commonplace in new physics scenarios and this constitutes our third template.
Finally, we omit tensor exchanges because no such well-motivated 
model leading to large $A_{FB}$ exists. We now discuss each template in turn.

\subsection{A flavour-nondiagonal $Z'$}
While additional $U(1)$ gauge symmetries are well-studied and very often
well-motivated, it is easy to see that most common variations would not
lead to a substantial $A_{FB}$. 
In models where the $Z'$ appears in the $s$-channel in 
the $q \bar q \to t \bar t$ process, the NP amplitude cannot 
have a non-zero interference with the QCD contributions. Consequently, 
obtaining the required $A_{FB}$ requires large $Z'$ couplings resulting 
in too large a correction to $\sigma(t \bar t)$. On the other hand, 
a flavour-changing $Z'$~\cite{Murayama} would appear in the 
$t$-channel in, say, $u \bar u \to t \bar t$. The corresponding contribution 
to $A_{FB}$ has two sources, viz. kinematic (due to the $t$-channel 
propagator) and, possibly a dynamic one as well (if the $Z'$ coupling 
is chiral). Analogous $\bar t c Z'$ or $\bar c u Z'$ couplings are 
disfavoured as these would set up flavour-changing neutral currents. 
Similarly, consistency with $B$-physics observables is simpler if 
the $Z'$ does not couple to the $b$-quark necessitating that 
the coupling to the top be right-chiral, thus leaving us 
with a Lagrangian of the form~\cite{Murayama}
\beq
{\cal L} \ni \gX \, Z'_{\mu} \, \bar u \, \gamma^{\mu} \, P_R \, t + h.c. \ .
\label{zpr:lag}
\eeq
The model also includes a small flavor-diagonal coupling to $u$ quarks
in order to avoid experimental constraints from like-sign top 
production data from the Tevatron~\cite{Murayama}. 
However, this as well as issues regarding anomaly cancellation are not
relevant to the discussion at hand and, hence, we shall ignore them. 

\subsection{Diquarks}
Particles carrying a baryon number of $\pm 2/3$ occur in many models,
most frequently in those concerned with grand
unification~\cite{Hewett:1988xc}.  While both scalars and vectors are
possible, the latter would, typically, have a mass as large as the
symmetry breaking scale. The scalars can be light, though, and may
couple to a pair of quarks. Indeed, an example of such couplings can
be found within the minimal supersymmetric standard model (MSSM) if
$R$-parity were not conserved. Once again, the presence of such
couplings would generate a non-zero $A_{FB}$, for both kinematic and,
if the coupling is chiral, dynamic reasons. The fermion assignments
within the SM ensures that the couplings are indeed chiral in nature.
Of the various different diquark fields that are possible, clearly the
one that couples a $u$ to a $\bar t$ can generate the maximal amount
of $A_{FB}$ for given mass and coupling strength. In other words, the
relevant piece of the Lagrangian is \cite{Tait}
\beq
{\cal L} \ni \Phi^a \, \bar t^c \, T^a \, (\yS + \yP \, \gamma_5) \, u + h.c., 
\label{diq:lag}
\eeq
where $T_a$ denotes the appropriate color-coupling structure. As
can be expected, the cross-sections for $\Phi^a$ transforming as a
$\mathbf{\bar 6}$ of $SU(3)_C$ differs from that for a $\mathbf{3}$
only in colour factors. We will not discuss the two situations
separately and will restrict to the color triplet
diquark\footnote{Ref.~\cite{Tait} also investigated scalars 
  (with vanishing baryon numbers) in $\mathbf{8}$ and $\mathbf{1}$
  representations of $SU(3)_C$ but those were not found to be 
  very successful in consistently reproducing the relevant experimental 
  data.}.  
While Ref.\cite{Tait} admits generic $\yS$ and $\yP$, it is easy to see
that consistency in the $b$-sector motivates the choice of right-chiral couplings, 
viz, $\yS = \yP$.

\subsection{Axigluons}
Originally motivated as residues of unifiable chiral color
models\cite{Frampton&Glashow}, wherein the high energy strong
interaction gauge group $SU(3)_L \otimes SU(3)_R$ is spontaneously
broken to the usual $SU(3)_C$, the axigluon $A_\mu^a$ was nothing
but the octet gauge boson of the broken symmetry having a purely axial
vector coupling---of strength $g_s$---with the SM fermions. This model
has been probed, at Tevatron, both in the dijet~\cite{CDF_dijet_2008}
as well as the $t \bar t$ channel~\cite{DC&RMG,Antunano:2007da,CDF_NP_in_ttbar} and
masses $\ltap 1 \tev$ have been ruled out. Although such a scenario
immediately predicts $A_{FB}$, it has the right sign for this quantity
only for small values of $m_A$, well below the Tevatron
limits~\cite{DC&RMG}.  For larger $m_A$, the sign of $A_{FB}$ flips.

Motivated by this, Ref.\cite{Frampton} considered a variation with
a different embedding of the color group into a $SU(3)_A \otimes SU(3)_B$. 
The gluon and axigluon are now admixtures of the $SU(3)_A$ and $SU(3)_B$ with a 
non-trivial mixing angle $\theta_A$. 
Anomaly cancellation requires a fourth generation of quarks.
The new axigluon is `flavor non-universal' as its couplings with 
the $3^{rd}$ and $4^{th}$ generation quarks are different 
from those with quarks of the first two generations, the essential 
trick being to reverse the sign of the axial coupling of the top-quark, 
thereby reversing the sign of $A_{FB}$. In the bargain, additional 
vector-like couplings are introduced for the axigluon, leading to 
\beq
{\cal L} \ni \bar\psi \, \gamma_{\mu} \,
         (g_V^x + g_A^x \, \gamma_5) \, T_a \, \psi \, A^{\mu}_a \ .
\label{axi:lag}
\eeq
While the vector coupling is generation universal 
($g_V^x = - g_s \, \cot2\theta_A$), the axial coupling is not, with 
$\gAq = - g_s \,{\rm cosec} 2\theta_A$ for the first two generations 
and $\gAt = + g_s \,{\rm cosec} 2\theta_A$ for the last two. 
Note that, unlike in cases of the diquark and the $Z'$, wherein 
the new physics contribution only appears in the sub-process 
$u\bar{u} \to t\bar{t}$, for the axigluon, all quark flavors
are involved.

\section{Analytical Issues}
\label{sec:analytic}

As we are interested in the polarization asymmetry, 
we must calculate the cross-sections for different final
state helicity combinations. While the details are presented 
in the Appendix, note that, 
in each case, the square of the amplitude 
can be decomposed as
\[
\vert {\cal M}(\lt,\lbt) \vert^2 = 
{\cal A} + \lt \lbt \, {\cal B} + (\lt-\lbt) \, {\cal C} \ .
\]
where $\lt,  \lbt = \pm 1$ are twice the helicities of the top (antitop).
For observables relating to a single polarization, such as that we are 
interested in, the other must be summed over. 

It is easy to see that while the total $t\bar t$ cross-section and
$A_{FB}$ receive contributions from only from ${\cal A}$, the
polarization asymmetry $A_P \sim {\cal C}/{\cal A}$.  
It is, thus, worthwhile to examine the dependence of these terms
on the coupling constants. As far as the diquark is concerned (see
eq. \ref{diq:mesq}), the terms ${\cal A}$ (and ${\cal B}$) are even in
both the (pseudo)-scalar couplings $y_S$ ($y_P$), and thus, the
ensuing $A_{FB}$ is insensitive to their ($y_{S,P}$) sign 
(and, hence, to the chirality structure of the theory). 
The term ${\cal C}$ being proportional to the product $y_S \, y_P$, 
the polarization asymmetry picks up the chirality structure
unerringly. And while eq.\ref{zpr:mesq}, does not reflect this explicitly 
(owing to the fact that a $V+A$ structure was chosen), the story is similar 
for the $Z'$.

For the axigluon, the situation is bit more complicated. 
There exist pieces in ${\cal A}$ (and ${\cal   B}$), that are odd 
in the individual axigluon couplings, but being parity odd, most 
of them do not contribute to $\sigma_{t \bar t}$. The only such piece 
that contributes to $\sigma_{t \bar t}$ does so only away from 
the resonance---with the sign reversing as one crosses it---resulting 
in a subdominant contribution. Of course, $A_{FB}$ is very 
sensitive to the relative signs of the couplings in question, 
which was the original motivation of this model vis-a-vis the 
flavour-universal axigluon. Once again, the {\em sign} of $A_P$ 
would provide additional information about the aforementioned 
relative signs, thereby serving to establish the actual structure of the 
embedding of $SU(3)_C$ in the larger gauge group. 

Finally, note that, for each of the three cases, the piece ${\cal C}$ has 
a different angular dependence, including a parity-odd piece.
This leads one to hope that the construction of a rapidity-dependent 
$A_P$ would lend additional resolving power.

\section{Numerical Results}
\label{sec:numeric}

From the discussions in the preceding section, one expects,
in general, that each of the three models would be associated
with a different correlation between the three observables 
$\sigma$, $A_{FB}$ and $A_P$. 
However, since the particular values of these quantities 
depend sensitively on the parameters of the
model (i.e. the boson masses and couplings), it is conceivable that
the models could lead to similar values for these observables for
some unrelated points in the parameter space. Since such a degeneracy
would lead to the models being indistinguishable from each other, at least 
as far as such observables are concerned, we begin by delineating the 
situations where this might occur. 

To this end, we scan the parameter space for each of these models with 
the restriction that the couplings be 
perturbative\footnote{We impose this condition only until the 10 TeV scale. 
  Were we to demand that the theory remain perturbative until a 
  significantly higher scale, the parameter space would be further 
  restricted. However, in the absence of any knowledge of the
  ultraviolet completion, we desist from this.}.  
For the $Z'$, this implies $g_X < 2\pi$. For the diquark theory,
(to ensure that there are no relative normalization factors between the 
couplings in the different models under consideration) the same limit 
is imposed on $g_y$ = $\sqrt{2}y$ where, $y \equiv (y_S^2 + y_P^2)^{1/2}$.
In the case of axigluons, the perturbativity condition is imposed on the 
couplings $g_A$ and $g_B$ (associated with $SU(3)_A$ and $SU(3)_B$, respectively)
with both $g_A$, $g_B$ $< 2\pi$. This, in turn, translates to 
$10$\textdegree $< \theta_A < 45$\textdegree\cite{Frampton}.

As far as masses are concerned, the lower limits are model-specific.
It is easy to appreciate though that, for very large masses, all 
the models would, essentially, be equivalent to contact interactions. 
For very large masses, engendering a large enough $A_{FB}$ would, then,
need very large couplings, thereby coming into conflict with the 
perturbativity requirement discussed above. In effect, only 
$M_{new} \ltap $ 3000 GeV is relevant. 
Furthermore, we impose direct limits by taking into account
various considerations as detailed below.

In case of diquarks, one could argue for
$m_\Phi > 350 \gev$, motivated by the limits on squark
searches at the Tevatron\cite{Nakamura:2010zzi}. However, for a scalar
decaying primarily into a top (or a bottom) and a jet, these limits do
not apply and the corresponding limits are expected to be much weaker. 
Although such a dedicated study is yet to be reported, 
preliminary investigations~\cite{Choudhury:2005dg} suggest that 
$m_\Phi \gtap 200 \gev$ cannot yet be ruled out, and we shall adopt this. 

As for the $Z'$ model, much depends on the flavor structure of its coupling. 
If the $Z'$ is very light, then the branching fraction for $t \to Z' + u$
is no longer negligible. Further, if the $Z'$ decays hadronically, 
this would lead to an enhancement in the contribution to the $t \bar t$
production from the $lepton+jets$ channel as compared to the 
contribution from the $dilepton$ channel giving rise to deviations from 
the experimental measurements in the two channels. Ref.\cite{Murayama} has 
already considered this to find that $m_{Z'} \gtap 120 \gev$ is consistent
with current data and we shall adopt this limit as well. 

The limits on the flavor non-universal axigluon mass need more thought
though. Experimental searches, so far, have only considered 
flavor universal axigluons. Based on these, $m_A \ltap 1250 \gev$ is
ruled out by direct searches while the distortion in the $m_{tt}$
spectrum\cite{CDF_mtt} is too severe for $m_A \ltap 1400\gev$. 
However, it should be noted that the resonant production cross sections
for flavor non-universal axigluons differ on account of the differences
in their coupling to light quarks. Nonetheless, we do restrict to 
$m_A \gtap 1400\gev$. 

The parameter spaces thus defined, the numerical calculation is 
done using a parton level Monte Carlo routine. We use 
the CTEQ6L~\cite{CTEQ} parton distributions with the factorization scale 
set to $m_t$ with the later held to \mbox{172.5 GeV} to be 
consistent with the value used in 
measurements of cross-section~\cite{CDF_csec}
and $A_{FB}$~\cite{CDF_AFB_2010}. A $K$-factor~\cite{K-factor} of 1.33 is used to 
estimate the cross-section at NLO\footnote{In the absence of NLO calculations 
that incorporate the new physics effects under consideration, we use the SM K-Factor.}.

\begin{figure}[!htbp]
\centering
\subfigure[]
{
\includegraphics[width=3.1in,height=2.5in]{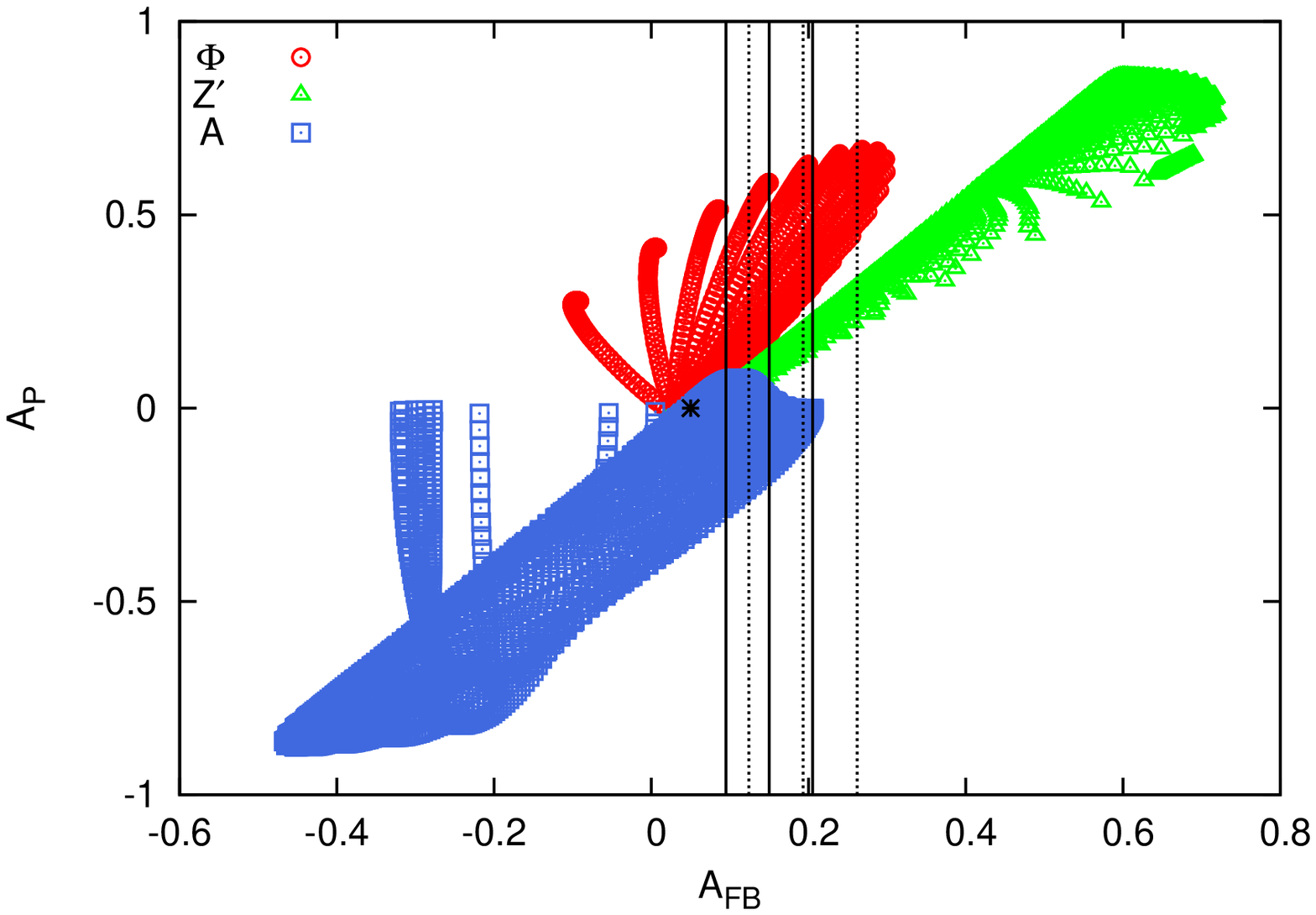}
\label{fig:scatter}
}
\subfigure[]
{
\includegraphics[width=3.1in,height=2.5in]{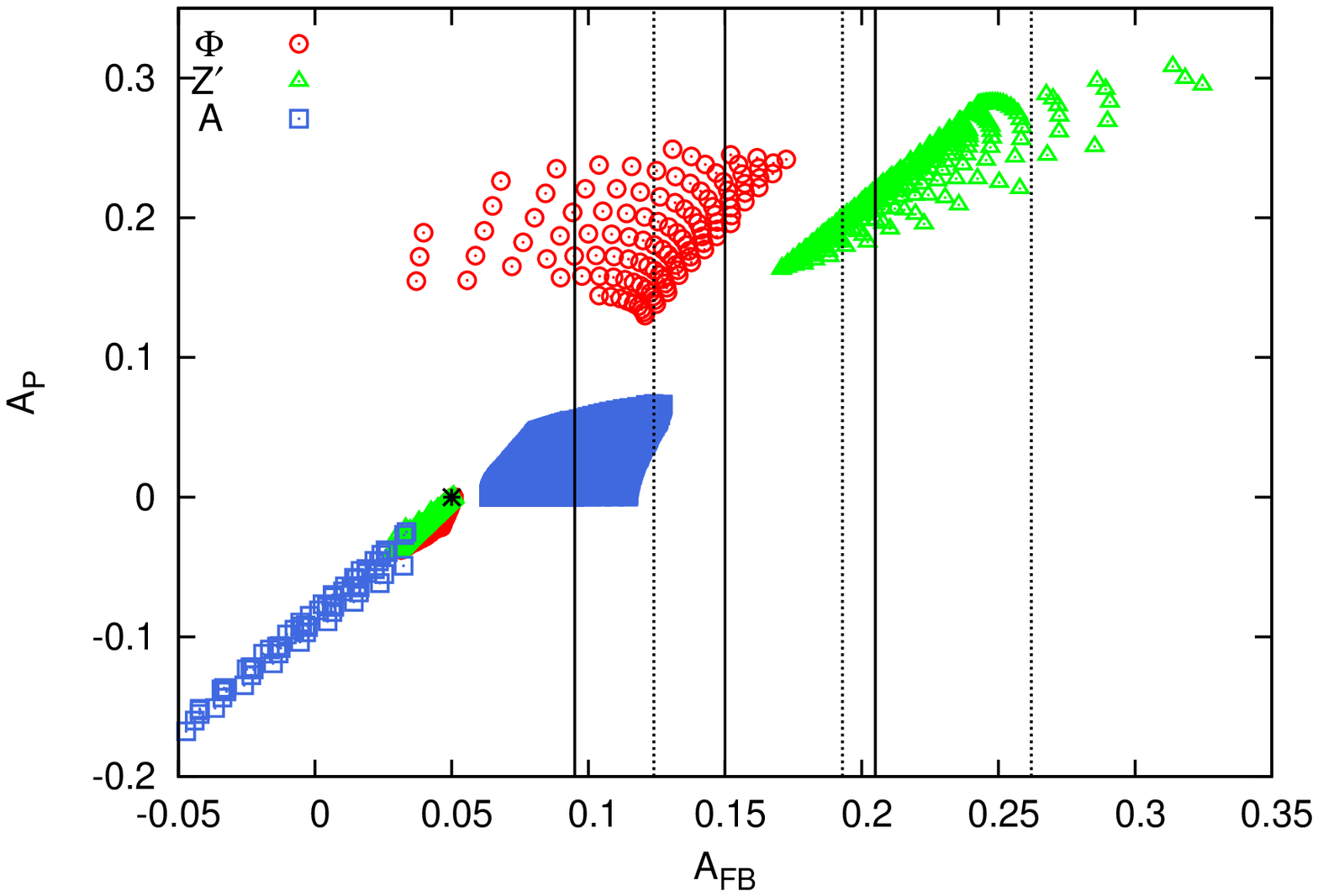}
\label{fig:sel_ref}
}
\caption{\em {\em (a)} Correlation between $A_P$ and $A_{FB}$ for 
different models. The vertical solid (dotted) 
lines correspond to the central value and 1-$\sigma$ bands of 
the new (old) CDF measurement of $A_{FB}$, namely 
\mbox{15.0$\%$ $\pm$ 5.5$\%$} \mbox{(19.3$\%$ $\pm$ 6.9$\%$)}. 
The `star' corresponds to the SM value at NLO. 
For each model, the scan is limited to only perturbative coupling strengths.
{\em (b)} As in the previous panel, but restricted to only those points 
that are consistent with the experimentally 
observed cross-section at the 1-$\sigma$ level and with restrictions
on $M_{\rm boson}$ as described in the text.
}
\label{fig:scatterplots}
\end{figure}

For each parameter-space point we calculate
$\sigma_{t \bar t}$, $A_{FB}$ and $A_P$ as defined above. Our
calculation of $A_{FB}$ includes the SM part~\cite{AFB_SMNLO}.  As
for $A_P$, the SM contribution is miniscule.  This is primarily
because a non-zero $A_P$ can arise only when the production
mechanism treats positive and negative helicity states
differently. Consequently, a pure QCD process can never give rise to
a non-zero $A_P$. In the SM, $A_P$ will arise only due to electroweak effects,
either at the tree level or through mixed EW-QCD higher order corrections,
and is, hence, rather small (${\cal O}(10^{-3})$). Given this, 
we restrict ourselves to the tree-level value of $A_P$ as indicated in the 
previous section.

In Fig.\ref{fig:scatter}, we show the correlation
between $A_P$ and $A_{FB}$ for each of the three models alongwith the 
experimentally allowed 1-$\sigma$ band for $A_{FB}$.  No
restrictions have been imposed on the physical observables, 
barring the aforementioned perturbative limit on the couplings. Note that the
splayed-finger like structure for $\Phi$ is but a reflection of the sampling
density in the masses and a finer sampling would have resulted in a
dense region in this plane. It immediately transpires that the three
models under discussion have very different correlations between the
observables. However, this is evidently misleading as a large faction
of the parameter space depicted here gives rise to a $A_{FB}$ well
outside the experimentally allowed range. This is even more so for
the total cross section. Clearly, large values of the asymmetries
would occur only for substantially large NP couplings and these,
typically, give rise to too large a cross section.

Imposing the restriction that the total $t \bar t$ production cross
section should lie within $1 \sigma$ of the experimentally observed
value and restricting $M_{boson}$ as described earlier, eliminates
bulk of the parameter points and leads to the correlation plot of 
Fig.\ref{fig:sel_ref}. 
One feature immediately stands out. 
For each of the models, the allowed regions split into two, one close to the SM
point and the other distinctly separate. 

\begin{figure}[!htbp]
\centering
\subfigure[]
{
\includegraphics[width=3.1in,height=2.5in]{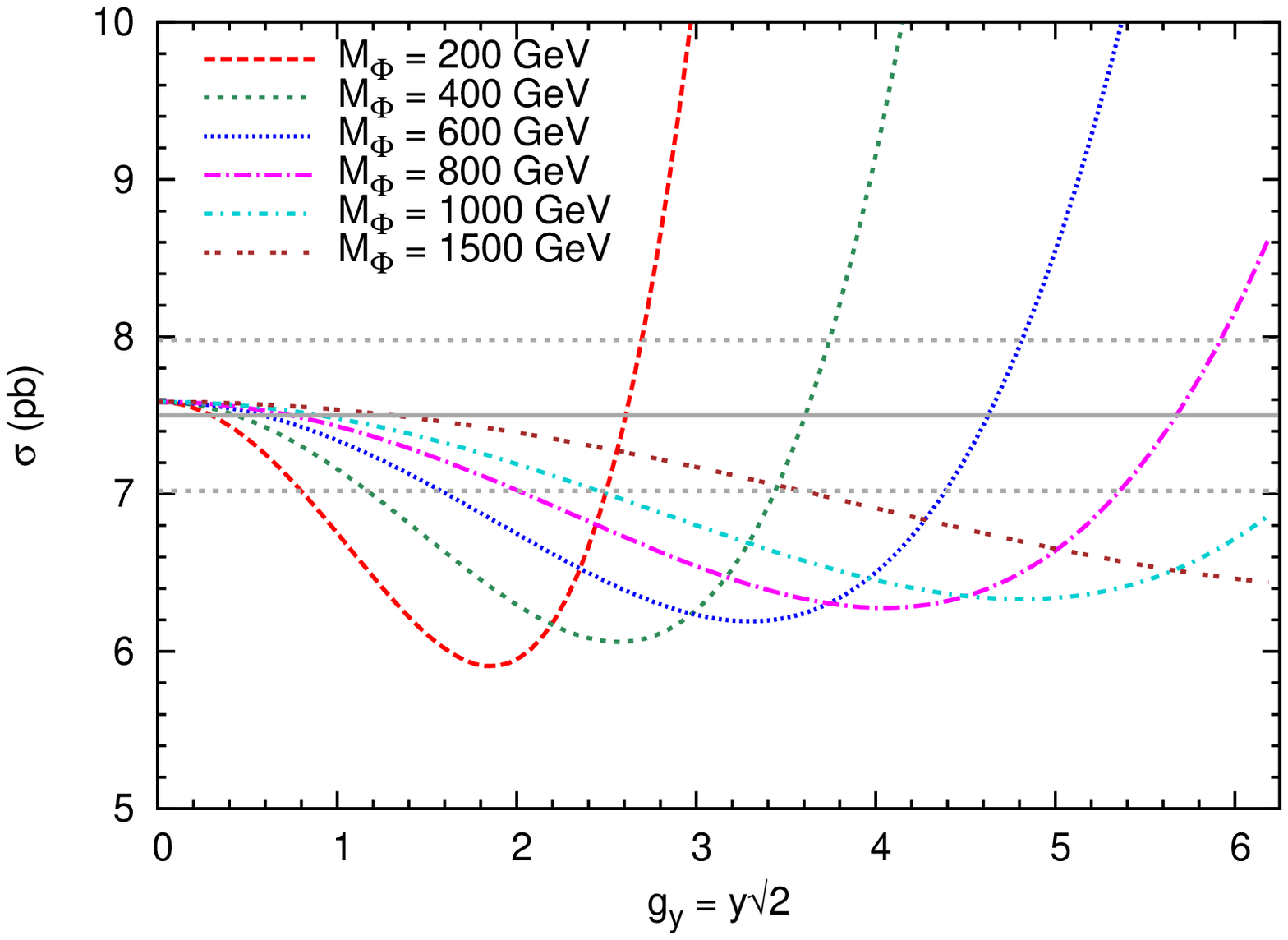}
\label{fig:dq_csecvscoup}
}
\subfigure[]
{
\includegraphics[width=3.1in,height=2.5in]{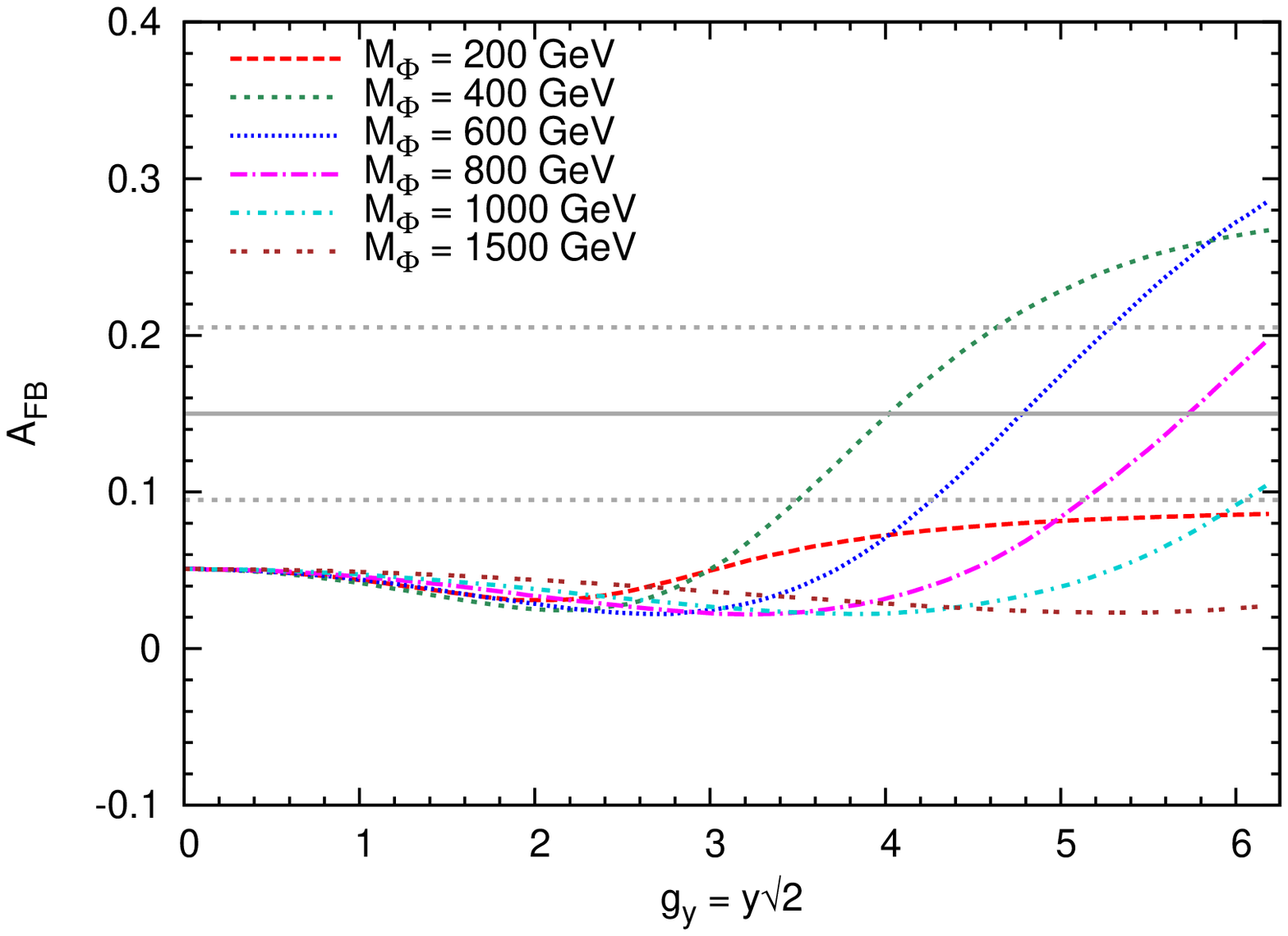}
\label{fig:dq_afbvscoup}
}
\caption{\em Variation in $\sigma(t \bar t)$ and $A_{FB}$ with coupling 
for different choices of $M_{\Phi}$.}
\label{fig:dq_vscoup}
\end{figure}

For the $Z'$ and the diquark, the reason for this is easy
to understand. Clearly, for small values of the NP coupling, the
cross-section is close to that within the SM and, hence, consistent
with the measured value.  However, for such small couplings, the
$A_{FB}$ generated is also small and not in consonance with the
experimental observation.  For a given boson mass, as one considers
larger values of the coupling, sufficient $A_{FB}$ can be produced
while the cross-section is prevented from becoming too large because
of the destructive interference between the SM and NP pieces. In fact, 
for an intermediate range of the coupling, the cross section falls 
well {\em below} the SM value. The location and the extent 
of the range naturally depends on the mass of the boson.
For the case of the diquark, this is illustrated in Fig.\ref{fig:dq_vscoup}. 
The situation is analogous for the $Z'$. 

\begin{figure}[!htbp]
\centering
\subfigure[]
{
\includegraphics[width=3.1in,height=2.5in]{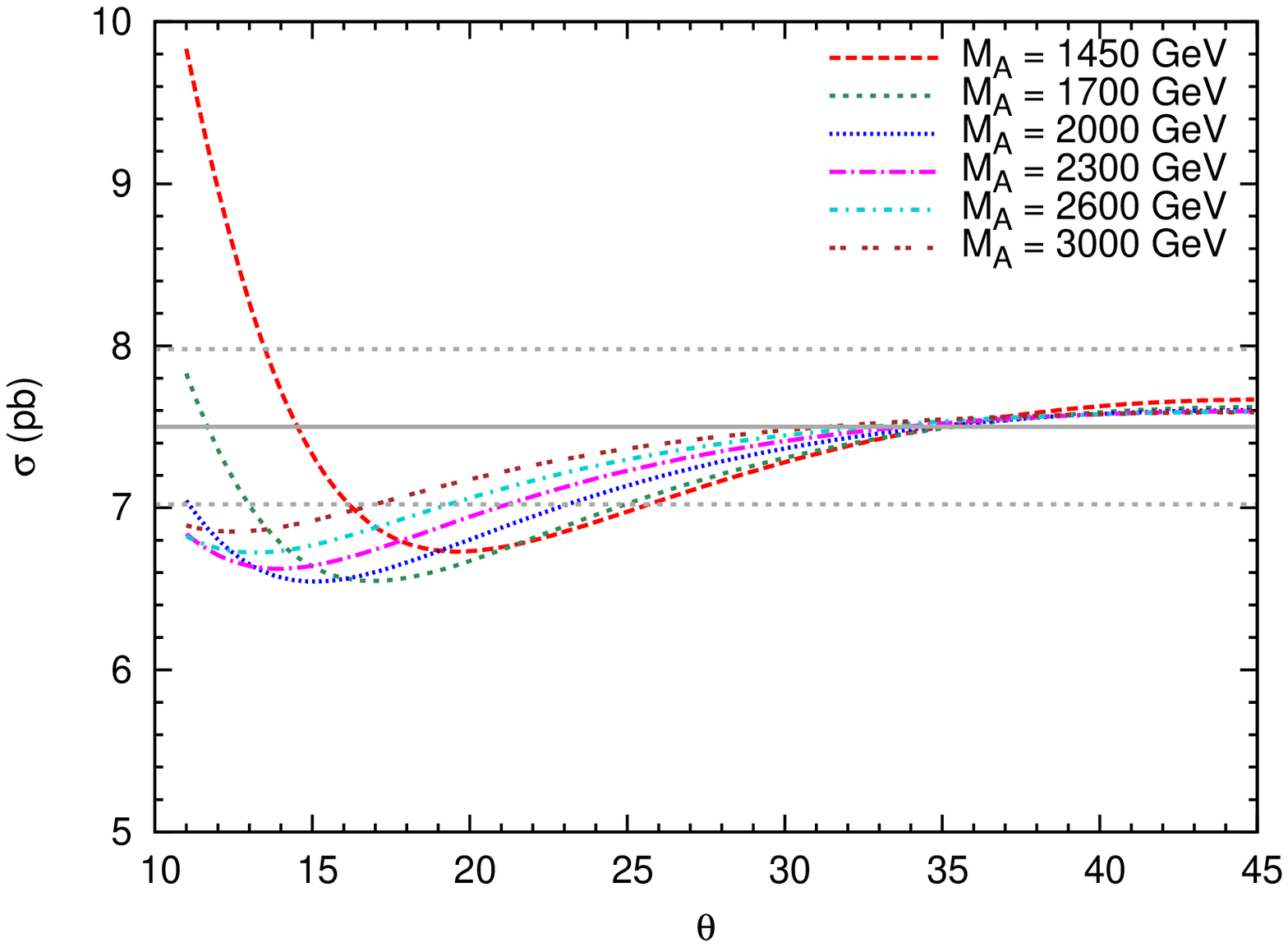}
\label{fig:ax_csecvscoup}
}
\subfigure[]
{
\includegraphics[width=3.1in,height=2.5in]{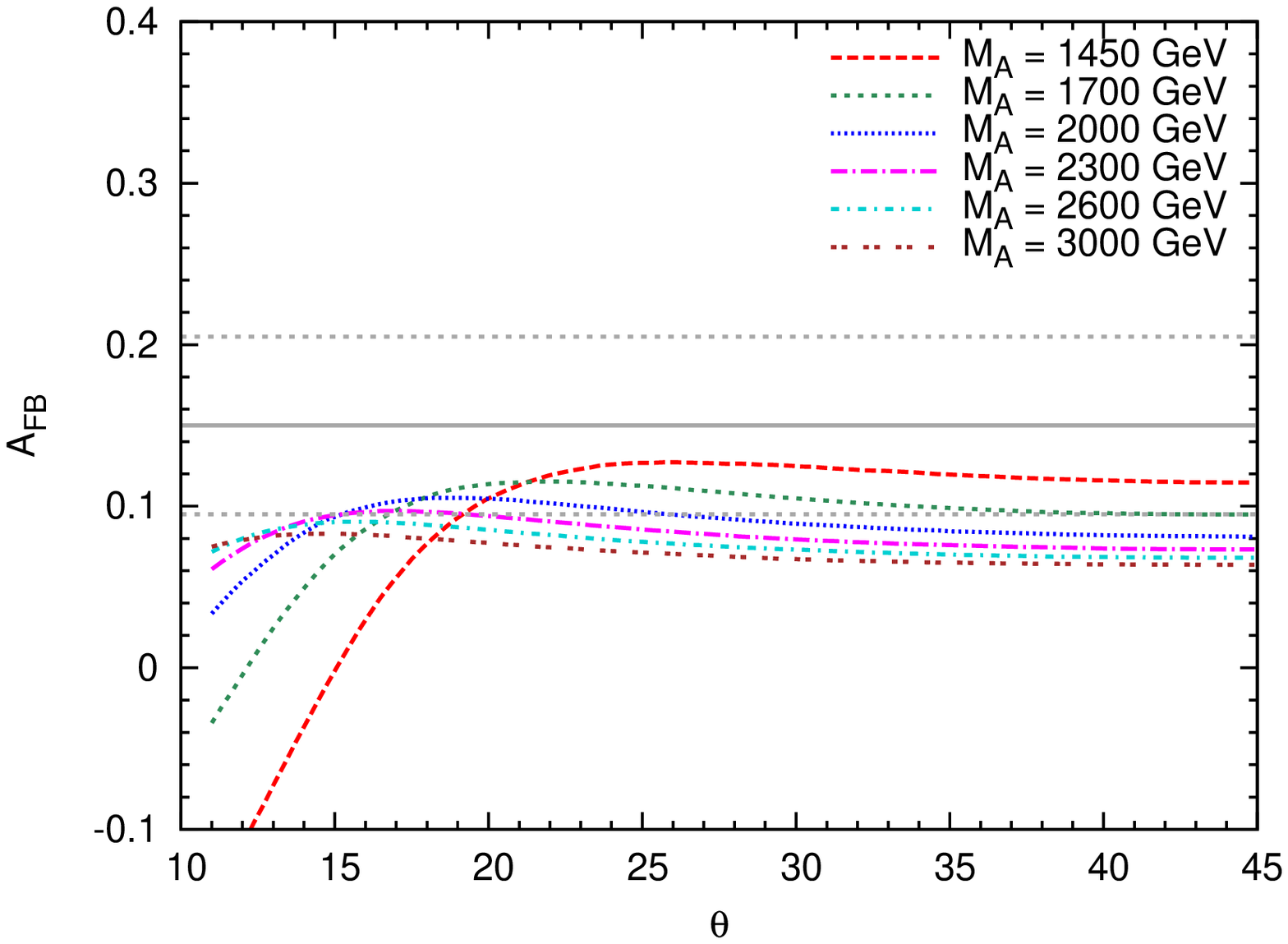}
\label{fig:ax_afbvscoup}
}
\caption{\em Variation in cross-section and $A_{FB}$ with $\theta$ 
for different choices of $M_A$.}
\label{fig:ax_vscoup}
\end{figure}

For the axigluon, the situation is somewhat more complicated. 
Here the interference may be constructive or destructive depending
upon the values of $\hat s$ and $M_A$ under consideration. 
For $M_A >$ 1400 GeV, the interference is mostly destructive though. 
For $\theta_A \approx$ 45\textdegree, the situation is identical to that
in the case of the flavor universal axigluon, with the exception that
now the relative minus sign between the $\gAq$ and $\gAt$ yields $A_{FB}$ that
is of the right sign as well as the right magnitude.
As $\theta_A$ decreases, the destructive interference takes over leading to 
$\sigma(t \bar t)$ values that are too low (see Fig.\ref{fig:ax_csecvscoup}).
For $M_A <$ 2000 GeV, a further increase in the strength of the couplings
allows the pure NP term to become dominant and the cross-section increases
once again to become consistent with data. 
Due to this, a part of the range of $\theta_A$ below 25\textdegree\ gets ruled
out. Again, the actual location and extent of the range depend on the 
particular value of $M_A$.
On the other hand, for $\theta_A \ltap$ 20\textdegree, $A_{FB}$ is only produced
in small amounts (Fig.\ref{fig:ax_afbvscoup}). The resultant picture that emerges
is the tail of points near the bottom left of Fig.\ref{fig:sel_ref} which are 
consistent with the measurement of $\sigma (t \bar t)$ but not with that of $A_{FB}$.

It is interesting to note that, in Fig.\ref{fig:sel_ref},
for the regions that are consistent with the measured $A_{FB}$ values
(whether the old measurement or the new one), there is no overlap between
the three models. It is true that, consistently flipping the
relative sign between the axial-- and vector--like
couplings\footnote{Of course, this would necessitate a different embedding
  in the gauge group.}, 
would flip the sign of $A_P$ while $\sigma$ and $A_{FB}$ remain unaltered. 
However, even after such a variation, the flavor non-universal axigluon case
would still retain its distance, in the $A_{FB}-A_P$ plane, from the
other two models. Thus, a measurement of $A_P$ should, in principle, be able
to distinguish the models even before the corresponding boson has been 
directly observed, this being particularly true of the axigluon.

Let us also note at this point that analyses of Ref.~\cite{Godbole:2010kr} indicate 
that polarization values of around $15\%$--$20 \%$ can be measured at the
Tevatron with $10$ fb$^{-1}$, with good significance. Further, in addition to
the azimuthal asymmetry explored in Ref.~\cite{Godbole:2010kr}, a simple
forward backward asymmetry of the decay lepton w.r.t. the beam direction, may 
also be used at the Tevatron and hence one would be able to increase the
sensitivity further.  From Fig.~\ref{fig:sel_ref} one sees thus that the
values of polarization predicted by the different models seem large enough
to afford measurement at the Tevatron. 

While discrimination of the axigluon model from the other two seems 
clear, that between the  diquark and the Z' models is not so straightforward.
For one, they lead to overlapping ranges in both $A_P$ and $A_{FB}$. 
Furthermore, Figs.\ref{fig:scatterplots} have been drawn with the error 
bars suppressed. Incorporating the latter would render the two apparently
disparate population zones overlapping. If, however, one were to flip the 
relative sign between the scalar and pseudoscalar couplings of the diquark, 
the dissimilarity between the models would become quite conspicuous. 

\begin{figure}[!htbp]
\centering
\subfigure[]
{
\includegraphics[width=3.1in,height=2.5in]{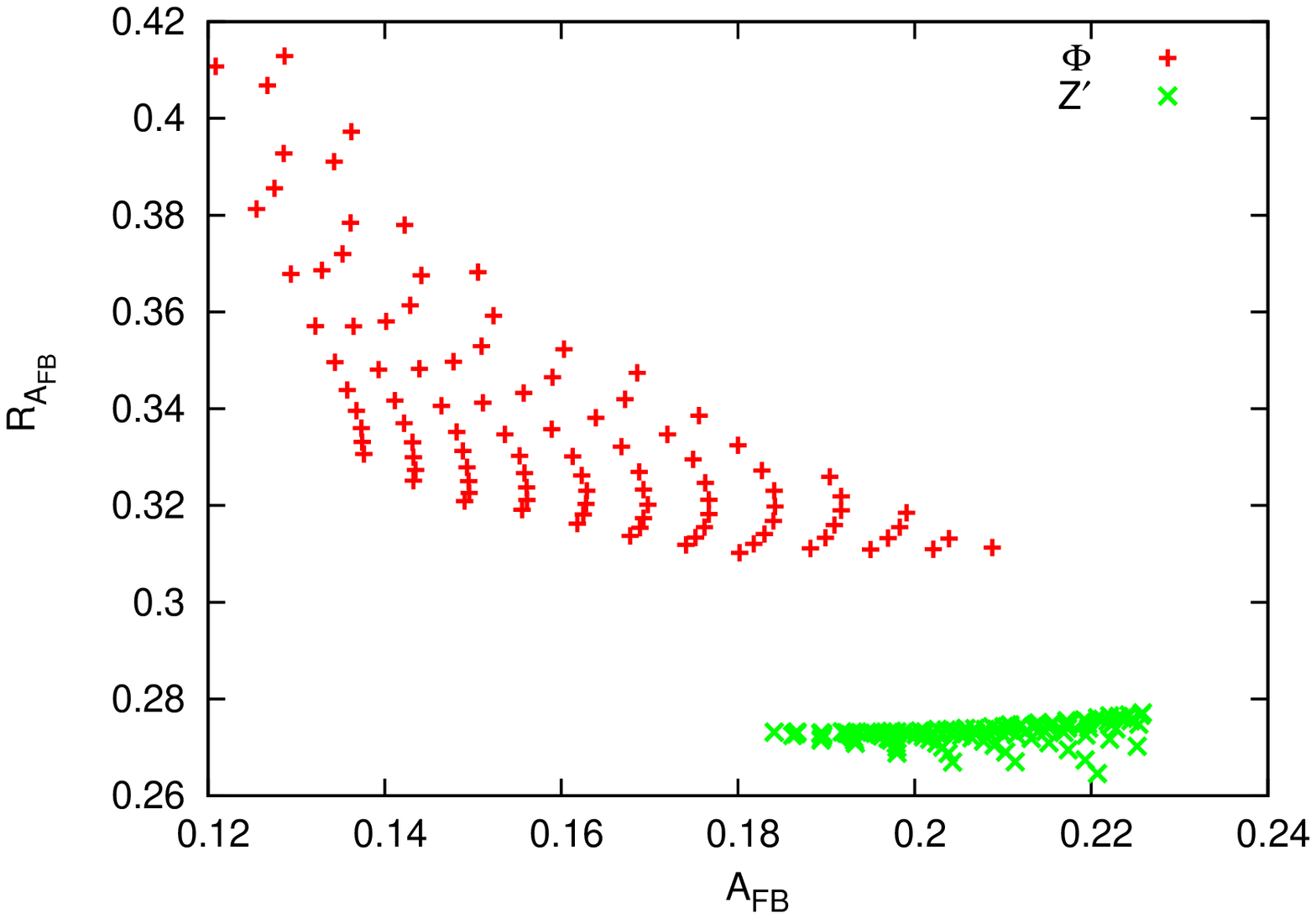}
\label{fig:AFB_ratios}
}
\subfigure[]
{
\includegraphics[width=3.1in,height=2.5in]{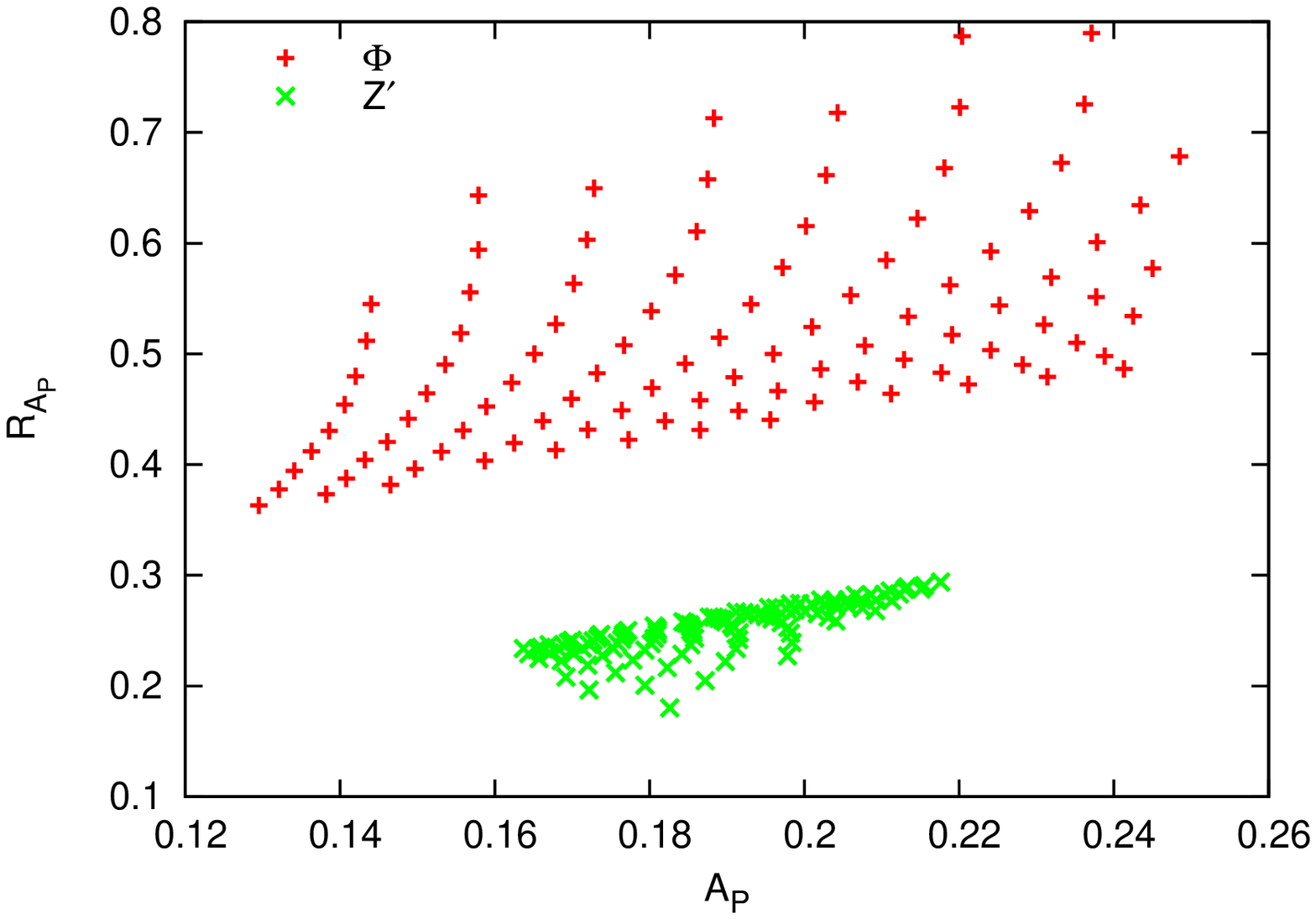}
\label{fig:AP_ratios}
}
\caption{\em $R_A$ vs $A$ for forward-backward asymmetry and
polarization asymmetry.  The points depicted are consistent with the
CDF measurementc of $\sigma(t \bar t)$ and $A_{FB}$ at the 1-$\sigma$
level. Also imposed are the perturbativity constraint on the couplings 
and the constraints on $M_{\rm boson}$ as described in the text.}
\label{fig:ratios}
\end{figure}

Nevertheless, even without changing the relative signs of $y_S$ and $y_P$
and with a view to making a more robust distinction between the two models, 
we propose observables that exploit the rapidity distributions 
in the two scenarios. With the typical values of the \mbox{$t$-channel} 
masses being different in the two scenarios (see Figs.2 \& 3),
it is but natural that the typical rapidities would be different. 
To this end, we define the ratios $R_{A_P}$ and $R_{A_{FB}}$ where
\begin{equation*}
R_A = \dfrac{A(|\Delta y|<1)}{A(|\Delta y|\geq1)} \ ,
\end{equation*}
and $\Delta y$ is the difference between the rapidities of the top and
the anti-top.
The CDF Collaboration does report a measurement of $A_{FB}$ in 
different $\Delta y$ regions \cite{CDF_AFB_2010}, leading 
to $R_{A_{FB}} = 0.043 \pm 0.194$ with the SM expectation being 
$R_{A_{FB}}(SM) = 0.317 \pm 0.053$. While this seems to offer a potentially 
strong discriminator, as we shall soon see, the associated errors
do not, yet, allow for discrimination between models.
Instead, we advocate an examination of correlations involving $R_{A_P}$ 
and $R_{A_{FB}}$. 

In Fig.\ref{fig:ratios}, $R_{A_{FB}}$ and $R_{A_P}$
are plotted against $A_{FB}$ and $A_P$ respectively. For
consistent comparison with CDF results~\cite{CDF_AFB_2010}, the
asymmetries are calculated in the $t \bar t$ rest frame and include
the NLO contribution from the SM~\cite{CDF_AFB_2010}. It can be seen
that the diquark and $Z'$ models populate different regions in the
$R_{A_{FB}}-A_{FB}$ and $R_{A_P}-A_P$ planes and the separation is
quite distinct.

Once again, Fig.~\ref{fig:ratios} does not
incorporate errors in the measurement of the quantities plotted. The
statistical part thereof can be easily estimated.
We obtain the efficiencies (i.e. including
kinematic cuts, event selection efficiencies etc.) from the number
of events reported in the CDF $A_{FB}$
measurement\cite{CDF_AFB_2010}, and assume that the corresponding
efficiencies for $A_P$ measurement would be similar. Calculating the 
expected number of events in this way, we then add the errors in quadrature.
While this might seem a trifle optimistic, we compensate by
restricting our projections to an integrated luminosity ($\sim$ 4.2 $fb^{-1}$)
less than what has already been analysed in Ref.\cite{CDF_AFB_2010}.

\begin{figure}[!htbp]
\centering
\subfigure[]
{
\includegraphics[width=3.1in,height=2.5in]{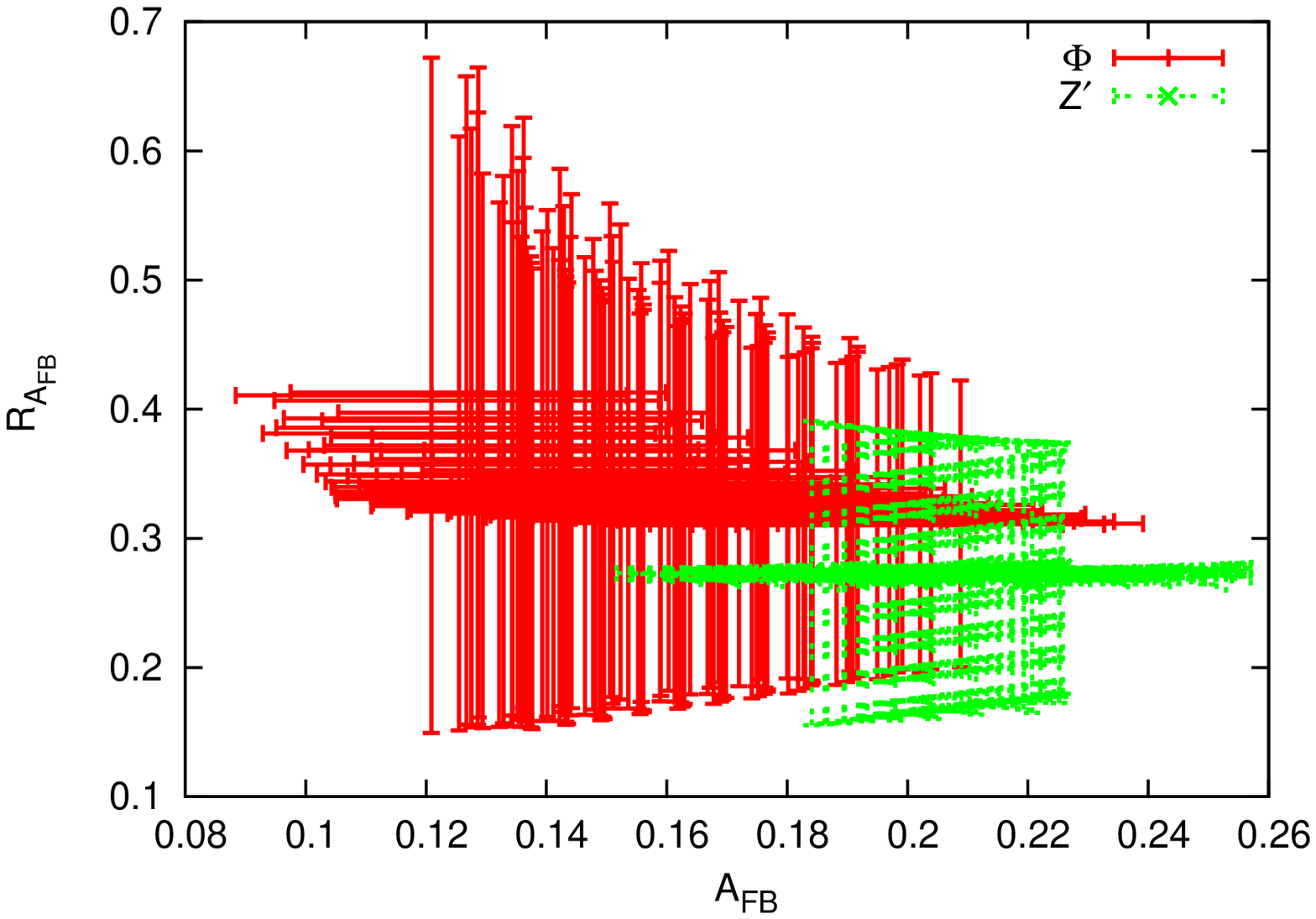}
\label{fig:AFB_ratios_w_err}
}
\subfigure[]
{
\includegraphics[width=3.1in,height=2.5in]{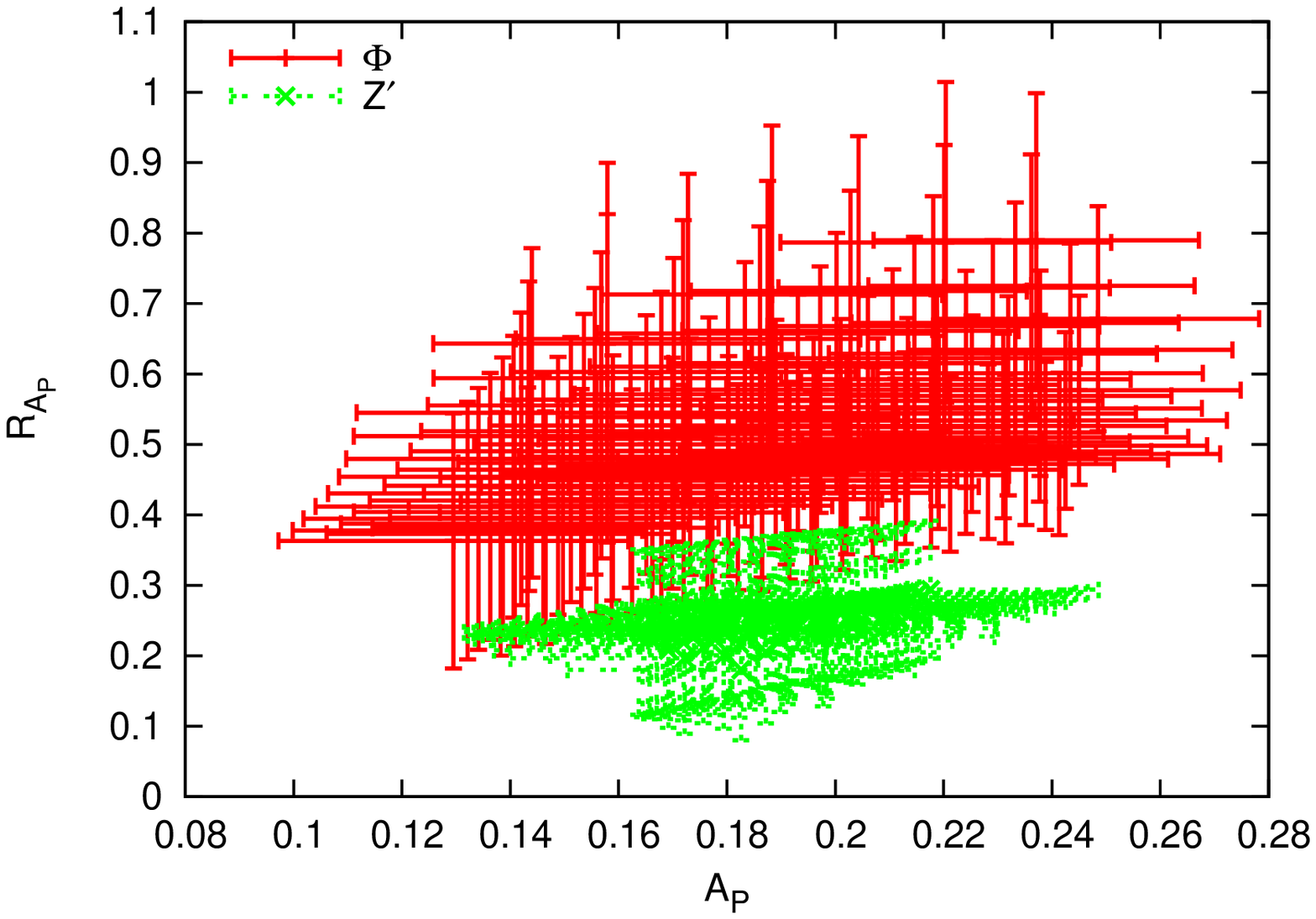}
\label{fig:AP_ratios_w_err}
}
\caption{\em Correlations involving $A_{FB}$, $R_{A_{FB}}$, $A_P$ and $R_{A_P}$ after 
inclusion of error bars. The points depicted are consistent with the same constraints
as explained in the caption of Fig.\ref{fig:ratios}.}
\label{fig:with_errors}
\end{figure}

\begin{figure}[!htbp]
\centering
\includegraphics[width=3.5in,height=2.6in]{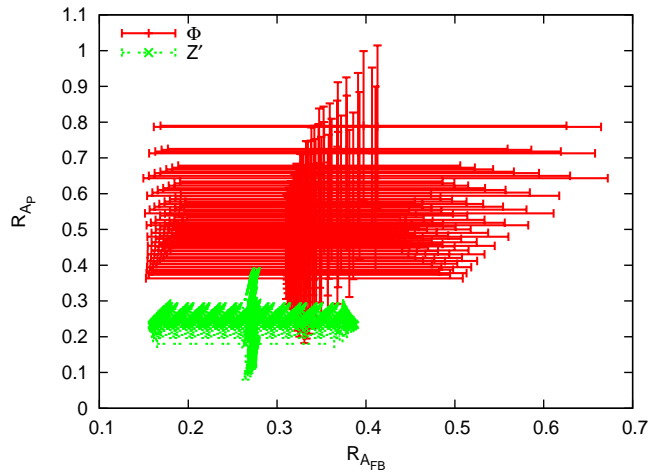}
\vspace*{-3ex}
\caption{\em Correlation between $R_{A_{FB}}$ and $R_{A_P}$.
The points depicted are consistent with the same constraints as explained 
in the caption of Fig.\ref{fig:ratios}.}
\label{fig:corr_ratios}
\end{figure}

In Fig.\ref{fig:with_errors}, we redraw Fig.\ref{fig:ratios} 
along with the error bars computed as described above.
Although the central values in the two scenarios are widely separated, 
a significant overlap between the error bars persists, thus
makes it difficult to distinguish between them. 
Indeed, the size of the errorbars also shows that the $R_{A_{FB}}$
measurement, on its own, is, as yet, incapable of conclusively distinguishing 
the models from even the SM.
On the other hand, the correlation between $R_{A_P}$ and $R_{A_{FB}}$,
(Fig.\ref{fig:corr_ratios}) looks more promising. While some overlap exists 
still, the extent of overlap is distinctly smaller. While a quantitative 
analysis confirms this, we desist from sharpening our conclusions 
at this stage. It should also be realized that 
an analysis more sophisticated than 
what we have performed and including all the data on tape, 
is quite likely to further reduce the relative sizes of the errors.

\begin{figure}[!htbp]
\centering
\includegraphics[width=3.5in,height=2.6in]{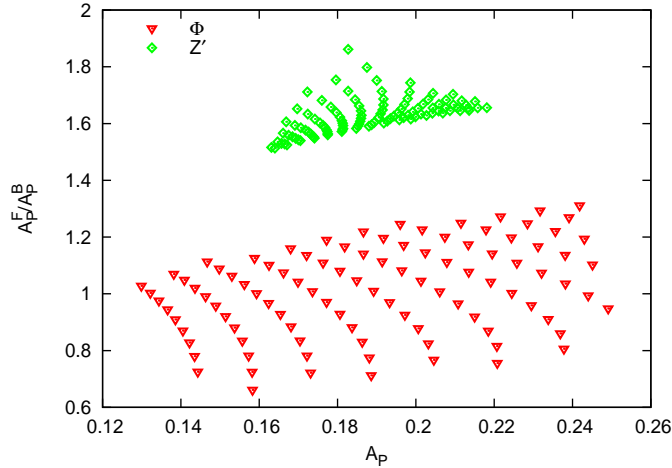}
\vspace*{-3ex}
\caption{\em Correlation between the ratio of $A_P$ in the forward and backward
hemispheres and total $A_P$. The points depicted are consistent with the same 
constraints as explained in the caption of Fig.\ref{fig:ratios}.}
\label{fig:AP_in_FnB}
\end{figure}

It should be realized that $R_{A_P}$ is not the only direction-dependent
asymmetry variable. Of the many such possible, we consider only one namely 
$A_P$ as calculated separately for the forward and backward hemispheres.
In particular, we plot, in Fig.\ref{fig:AP_in_FnB}, the ratio of $A_P$ 
in the two hemispheres against the total $A_P$.
Again, one finds that the diquark and $Z'$ models give rise to markedly 
different correlations. 

However, as the experience with $R_{A_P}$ has shown,
the effect of the separation is likely to be diluted by the errors involved.
With experimental errors under control, a combination of observables as 
described above, can be expected to distinguish between the ``explanations'' 
for the $A_{FB}$ ``anomaly'' quite successfully.

\section{Observables at the LHC}
\label{sec:LHC}
\begin{figure}[!htbp]
\centering
\includegraphics[width=3.5in,height=2.6in]{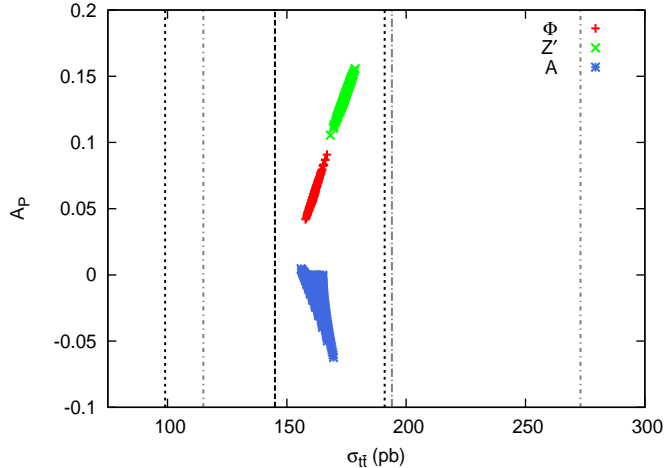}
\vspace*{-3ex}
\caption{\em Correlation between $A_P$ and $\sigma(t \bar t)$ at the LHC 
{\em ($\sqrt{s}$ = 7 TeV)}. The vertical lines show the 1-$\sigma$ interval 
of the CMS~\cite{CMS_ttbar} (dot-dashed) and ATLAS~\cite{ATLAS_ttbar} (dashed)
measurements of the $t \bar t$ cross-section.
The points depicted are also consistent with the constraints mentioned 
in the caption of Fig.\ref{fig:ratios}.}
\label{fig:at_LHC}
\end{figure}

The initial state being symmetric at the LHC implies that no simple
forward-backward asymmetry w.r.t the beam direction can be defined.
Hence, we limit ourselves to the correlation between $A_P$ and 
$\sigma(t \bar t)$\footnote{The K-Factor in this case is estimated
by taking the ratio of the NLO cross-section quoted in Ref.~\cite{CMS_ttbar}
and our own LO calculation resulting in K=1.8.}.
For the parameter space points that are consistent with the Tevatron
measurements of cross-section and $A_{FB}$ at the 1-$\sigma$ level, we
find that the three models correspond to different correlations (see
Fig.~\ref{fig:at_LHC}).  In going from the Tevatron to the LHC, the
relative contribution from the $gg \rightarrow t\bar t$ sub-process
(which does not contribute to such asymmetries) increases as compared
to the $q\bar q \rightarrow t\bar t$ subprocess.  The quark initiated
subprocess suffers a further suppression as anti-quark fluxes are
diminished at a $pp$ collider.  As a result, the magnitudes of $A_P$
at the LHC are lower than those observable at the
Tevatron. Nonetheless, it is interesting to see that the separation of
the expected values of $A_P$ into different islands is a trend that
persists. That the $A_P$ for the axigluon is now almost universally
negative (as oposed to the Tevatron where positive $A_P$ were allowed)
is but a consequence of the fact that a resonance is more easily hit
at the LHC.

It should also be borne in mind here  that the values  of polarization 
presented here correspond to those obtained by integrating  over the entire 
kinematical range.  It is clear that by making suitable cuts on the 
observables such as $m_{t \bar t}$ and/or $p_T^t$, the polarization may 
be  enhanced substantially, as has been seen in Ref.~\cite{Godbole:2010kr}.
It is also clear that the optimal cuts will be  again different for different
production mechanisms of the NP top pairs. 

\begin{figure}[!htbp]
\hspace*{-30pt}
\centering
{
\includegraphics[width=2.4in,height=2.8in]{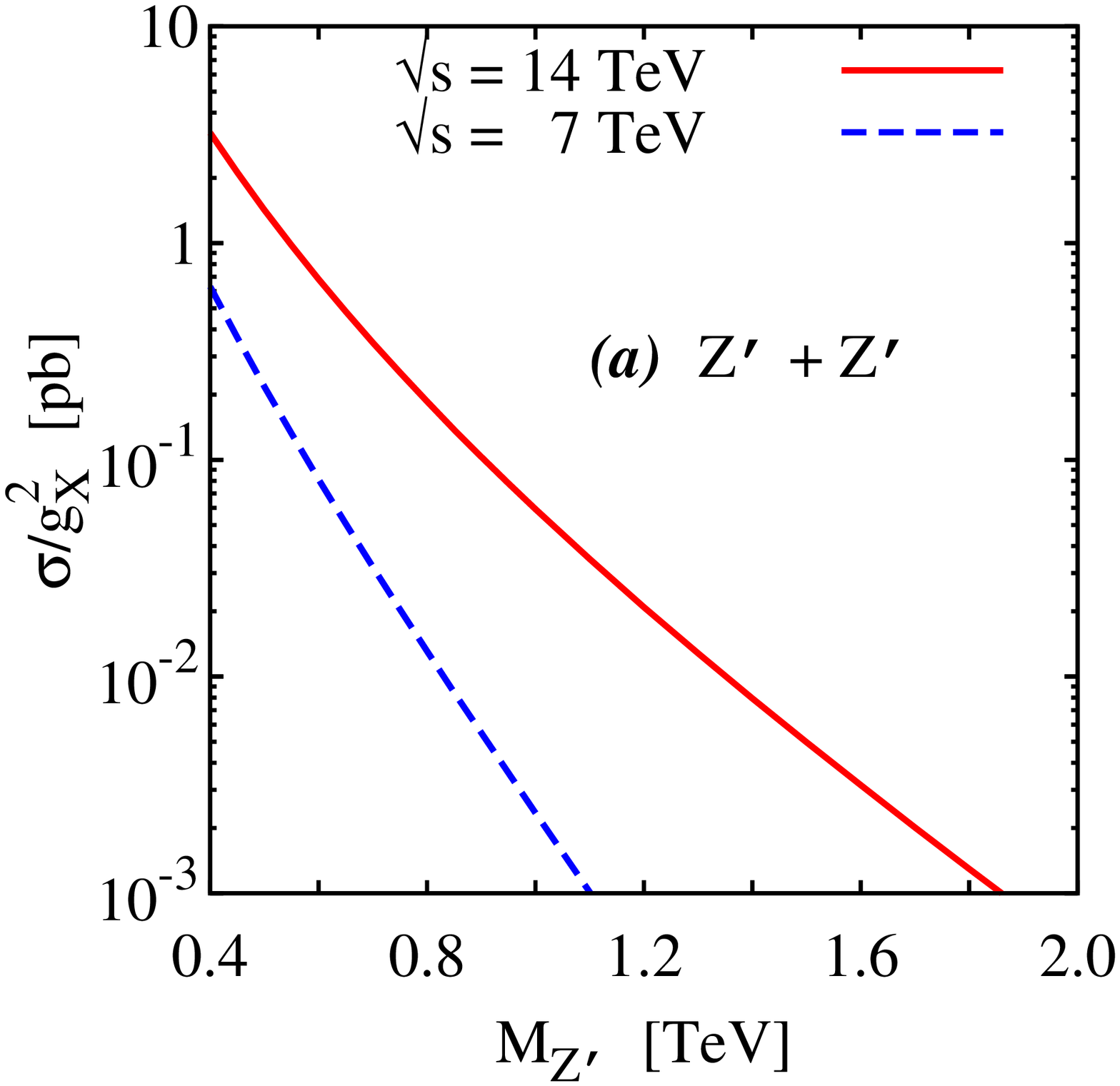}
}
\hspace*{-30pt}
{
\includegraphics[width=2.4in,height=2.8in]{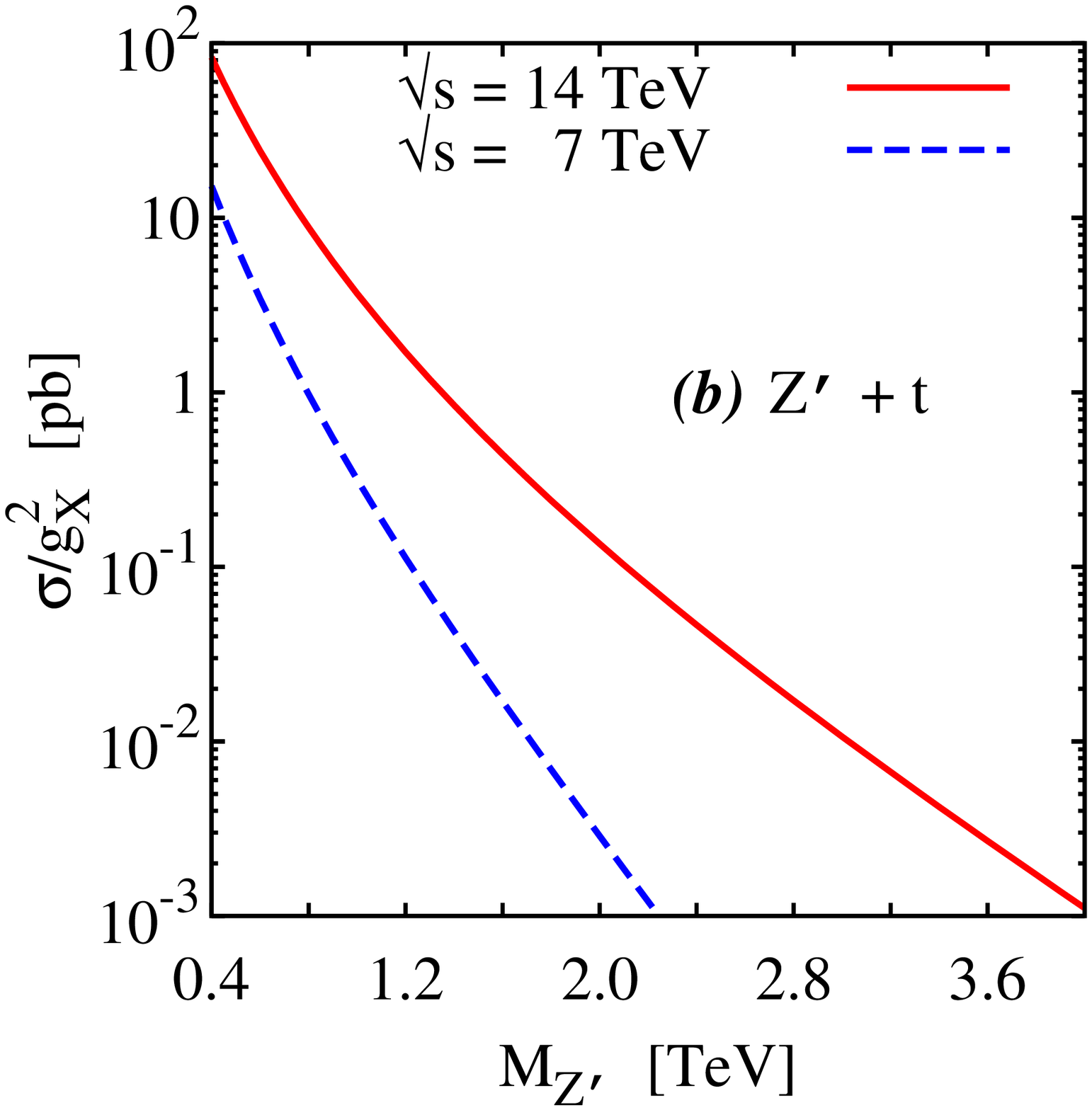}
}
\hspace*{-30pt}
{
\includegraphics[width=2.4in,height=2.8in]{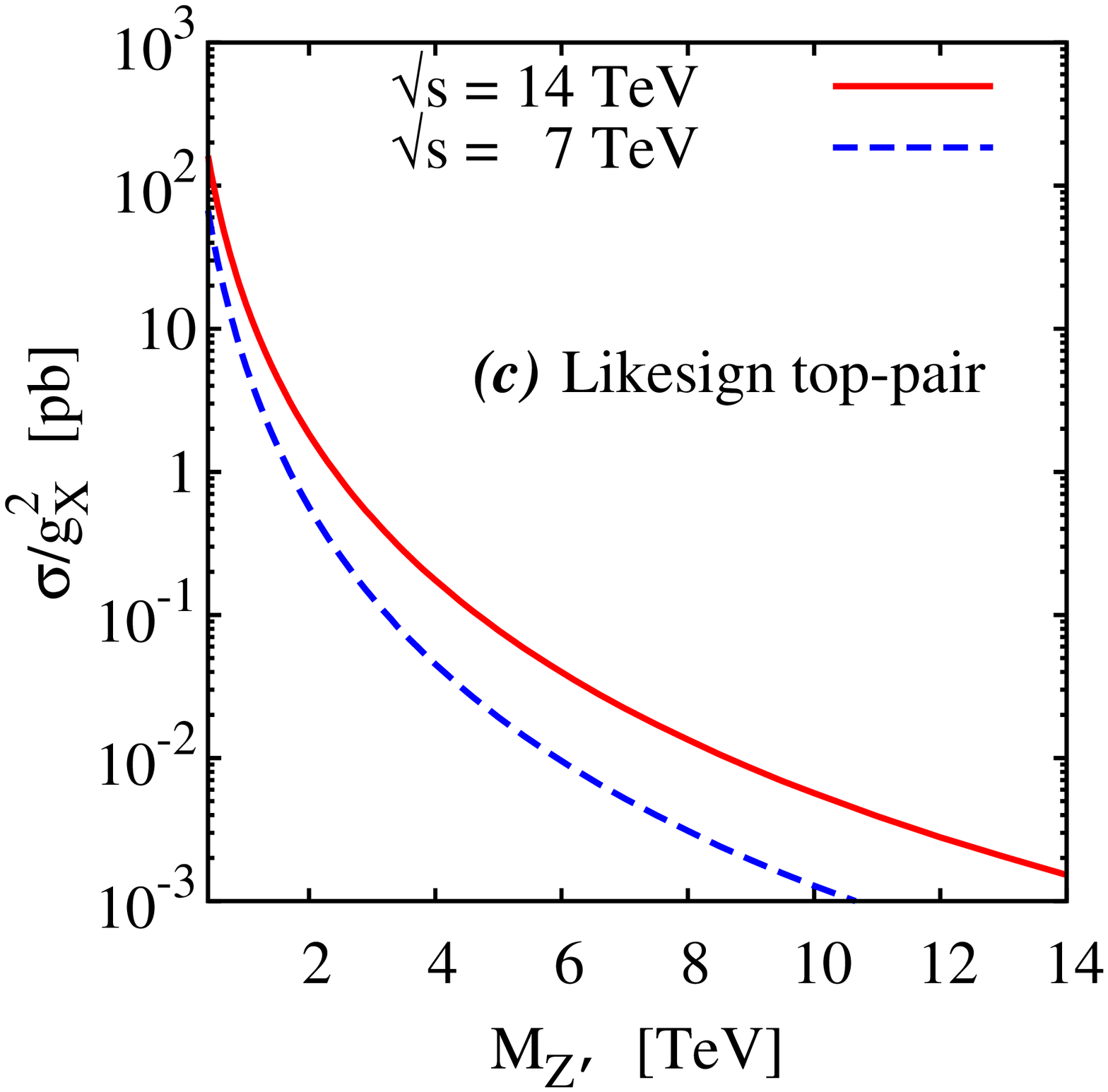}
}
\hspace*{30pt}
\vspace*{-30pt}
\caption{\em Several production cross sections at the LHC for the $Z'$ model. Note 
that obtaining the requisite $A_{FB}$ at the Tevatron requires the coupling 
to be typically larger ($g_X \gtrsim 1$).}
\label{fig:Zp_prod}
\end{figure}

At the LHC, there also exists the possibility of direct detection of 
these models through the production of the corresponding boson. 
For diquarks and axigluons (albeit, the flavor universal variety), 
the production cross-sections have been calculated in Refs.~\cite{Tait}
and~\cite{DC&RMG} respectively. For the $Z'$, direct detection is possible through
pair production of $Z'$s or production of a $Z'$ in association with a $t$.
We present the corresponding cross-sections for these processes in 
Fig.~\ref{fig:Zp_prod}, assuming that eq.(\ref{zpr:lag}) is the only term in the 
interaction Lagrangian. 

Although the cross-sections for the first two processes seem quite
large, claims about detectability should be made cautiously. Note that
such $Z'$ would decay into a top and a light quark (unless other
couplings are switched on). Thus, these production processes would,
essentially, lead to a $t \bar t$ pair accompanied by one or two hard
jets. The QCD background for the same is quite large. While invariant
mass reconstruction would, in principle, distinguish between signal
and background, in practice this is not an easy
task~\cite{Choudhury:2005dg}, especially with two top-quarks present.
This observations holds for diquark production as well. 

Relatively easier to look for is the production of a like-sign top-pair, 
for which this model predicts a large rate. Indeed, as Fig.~\ref{fig:Zp_prod}(c)
shows, with the projected luminosities at the LHC, such a $Z'$ could be 
measured upto several TeVs even for $g_X = 1$ (which, essentially, takes us to 
the limit of a contact interaction). It should be noted, though, that a very 
large $m_{Z'}$ necessitates a large $g_X$ for producing the correct $A_{FB}$. 
Such a scenario should have made its presence felt in the very same 
process at the Tevatron itself.

\section{Summary}
\label{sec:summary}

The anomalously large forward-backward asmymmetry in top-pair 
production observed at the Tevatron continues to puzzle. Several 
models have been proposed to ``explain'' this. While some of them 
have their roots in well-motivated scenarios originally proposed 
to address other issues, some of the others are purely phenomenological 
in nature. The very fact that many different models of new physics 
can explain this anomaly renders difficult the identification of the 
best solution. 

In this note we have analysed the role that the longitudinal top
polarization ($A_P$) can play in discriminating between such models.
The scenarios proposed, typically, differ from each other in their
chiral structure and in the relative amount of FB asymmetry that is
generated by the dynamics as opposed to kinematic effects.  $A_P$,
being a pure parity violating effect, probes the former aspect.

To illustrate this point we choose from among the different proposed
models, three templates and calculate the predictions for $A_P$ in
each. As expected, we find different correlations between $A_P$ and
$A_{FB}$ in the three cases and some scenarios can be clearly
distinguished from others purely on this account.  However, bearing in
mind that experimental errors may reduce this apparent separation
between the models, we further construct the ratios $R_A \equiv A
(|\Delta y| < 1) / A (|\Delta y| > 1)$, with $\Delta y$ being the
rapidity separation between the top pair.  Correlation between
$R_{A_P}-R_{A_{FB}}$ seems quite promising in its ability to
separate the different models even after the inclusion of errors.
Given that such measurements are already being made at the Tevatron,
we hope that the adoption of our algorithm would serve to solve this
vexing issue.

We also look at the case of the LHC briefly. Again, a scan over the
parameter ranges of the three template models, consistent with the
Tevatron data, shows that they occupy three disjoint islands in the
$A_P-\sigma_{t \bar t}$ space.  Although the magnitude of $A_P$ is
smaller, owing to the increased importance of the $gg \to t \bar t$
subprocess, the situation could be improved further by the adoption of
apropriate kinematic restrictions.  In this particular exploratory exercise, 
as far as the LHC is concerned, we refrain from carrying out such studies.

While most of the models to explain $A_{FB}$ had been chosen by the
respective authors, so as to evade current direct search bounds, this
would not be the state of the matter in the LHC era. As we have
demonstrated, the production cross-sections for new particles inherent
to such models, are likely to be large at the LHC. Given the peculiar
couplings of such particles, commensurate search strategies need to be
devised, but, by no means, is this an insurmountable problem. Indeed,
even if the new particles were all heavy, the effective Lagrangian
responsible for generating the $A_{FB}$ would, typically, leave its
mark on SM processes at the LHC (see Sec.\ref{sec:LHC}).

It should be noted that, by their very nature, most of the scenarios
proposed are phenomenogical in nature and have concentrated on
explaining $A_{FB}$ while taking care of consistency with only the
most obvious of other observables. Indeed, a global fit would render
many of them inconsistent. For example, the diquark model would take
$R_\ell \equiv \Gamma(Z \to {\rm hadrons})/\Gamma(Z \to \ell^-
\ell^+)$ away from the LEP/SLC measurements~\cite{Bhattacharyya:1995bw}. 
As for the $Z'$ model, the couplings and masses required to give a good
fit to both $\sigma_{t \bar t}$ and the new measurements of $A_{FB}$
are disfavoured by the lack of observation of like-sign top pairs at the
Tevatron~\cite{Murayama}. The flavor non-universal axigluon, on the
other hand, is disfavoured by $B$-physics observables~\cite{Chivukula:2010fk}.
While such problems might seem specific to these models, 
variations of these and others such are almost endemic to all such endeavours.
They can, of course, be cured by introducing compensatory effects in the
respective models. Thus, rather than rejecting them on account of such
disagreements, direct searches are the best bet. 

It is here that algorithms such as those we propose are particularly 
useful. Apart from discriminating betweeen models, such analyses 
can also serve to form an information basis on which more sophisticated 
model-building can be based. 

\textit{\textbf{Note Added :}}
As we were finalising the manuscript, two papers,~\cite{Cao:2010nw} and
~\cite{Jung:2010yn}, have appeared. In the former, the authors have studied 
different polarization observables, one of them being the polarization of 
the $t$ quark that we consider. While they look at different models 
proposed to explain the $A_{FB}$, they concentrate solely on the LHC (at $7$ TeV 
as well as $14$ TeV) and do not consider correlation with $A_{FB}$. 
Ref.~\cite{Jung:2010yn}, on the other hand, looks at the effects at the Tevatron
but it is an extension of model-independent analysis of $A_{FB}$ done in their
earlier paper~\cite{Jung:2009pi} to predictions for correlations of 
$t$ and $\bar t$ polarizations. Since CP conservation relates $t$ and $\bar t$
polarizations, we consider it sufficient to study the polarization of either 
one of $t$ or $\bar t$, which gives us the advantage of larger statistics.
Further, the analysis of Ref.~\cite{Jung:2009pi} is valid only when masses of 
exchanged particles are much higher than the cut off scale and hence there 
is no direct, easy comparison between the two approaches.

\begin{figure}[!htbp]
\centering
\includegraphics[width=4.5in,height=3.5in]{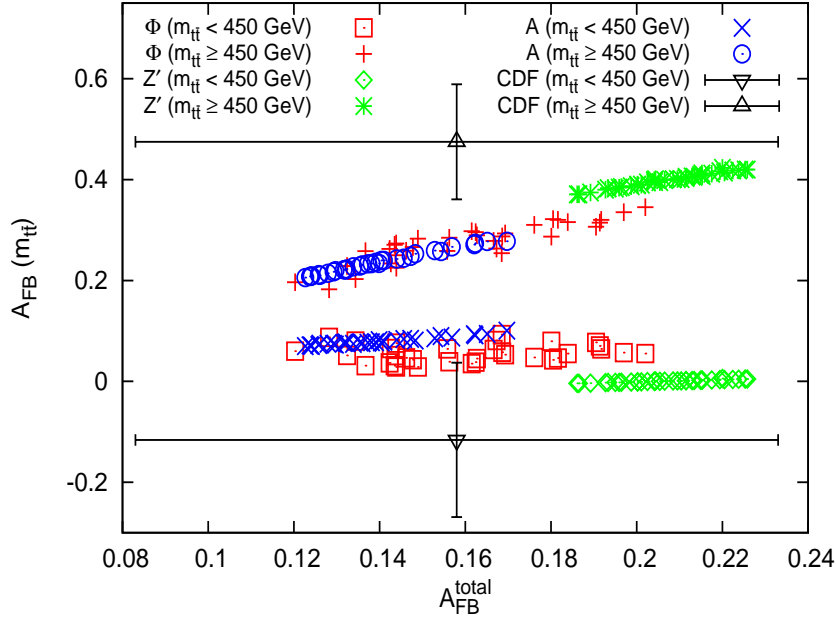}
\vspace*{-3ex}
\caption{\em $A_{FB}$ in the regions $m_{t \bar t} < 450$ GeV and $m_{t \bar t} \ge 450$ GeV
compared to the total $A_{FB}$. $A_{FB}$ is calculated in the $t \bar t$ rest frame.
The points depicted are consistent with the same constraints as explained 
in the caption of Fig.\ref{fig:ratios}.}
\label{fig:AFB_in_mtt}
\end{figure}

After the submission of our manuscript, the CDF
Collaboration published a new analysis of
$A_{FB}$~\cite{Aaltonen:2011kc} including its dependence on the mass
and rapidity of the $t \bar t$ pair.  The analysis in different
rapidity regions is identical to that in Ref.~\cite{CDF_AFB_2010}
which we have already considered.  Further,
Ref.~\cite{Aaltonen:2011kc} reported that $A_{FB}$ is found to be
larger in the high invariant mass region ($m_{t \bar t} \ge 450$ GeV) 
than in the low invariant mass region ($m_{t \bar t} < 450$ GeV).  
While Ref.~\cite{Aaltonen:2011kc} itself pointed out that the
analysis is still fraught with theoretical uncertainties, we carried
out a preliminary investigation which showed that this feature is
borne out by all the model templates that we have studied
(Fig.\ref{fig:AFB_in_mtt}) and does not offer any additional
resolution between the models under consideration.
In fact, this was to be expected in view 
of the facts that ({\em a}) the total cross section, which is dominated 
by the small $m_{t \bar t}$ range agrees very well with the SM expectations
and ({\em b}) the measured $A_{FB}$ is substantial.
Furthermore, even if uncertainties, both theoretical
and experimental, can be reduced, the crucial fact is that, unlike
$A_P$, this observable has little sensitivity to the chiral nature of the new
physics couplings.

\begin{center}\begin{small}\textbf{ACKNOWLEDGEMENT}\end{small}\end{center}
R.G. and S.D.R. wish to acknowledge support from the Department of Science and
Technology, India under the J.C. Bose Fellowship scheme under grant nos. SR/S2/JCB-64/2007
and SR/S2/JCB-42/2009 respectively. P.S. would like to thank CSIR, India for assistance 
under SRF Grant 09/045(0736)/2008-EMR-I.
This work was initiated during the 
\textit{$XI^{th}$ Workshop on High Energy Physics Phenomenology}
held at PRL, Ahmedabad. R.G. and P.S. would like to thank the organisers 
of the Workshop and the coordinators of the LHC Physics Working Group
for their hospitality and cooperation.

\newpage
\section*{\begin{center}Appendix\end{center}}
\renewcommand{\theequation}{A.\arabic{equation}}
We present here the squared matrix elements for the process 
$q \, \bar q \to t \, \bar t$, where, in the subprocess center 
of mass frame, $t$ has a velocity $\beta$ and subtends an angle 
$\theta$ with $q$. The final state quarks carry 
polarizations $\lt$ and $\lbt$ respectively. These can then be 
used to define the requisite density matrices. 
For each case, one can write
\beq
|{\cal M}|^2 = |{\cal M}_{SM}|^2 + |{\cal M}|^2_{NP}
\eeq
where the first part on the r.h.s. represents the SM 
contribution while the second represents the additional 
contribution due to the new physics (including the interference 
with the SM amplitude, and hence not positive definite). 
For the SM piece, we have, 
\beq
|{\cal M}_{SM}|^2 =
\dfrac{g_s^4}{18}\Big[ (1-\lt\lbt)(1 + c^2_\theta) 
+ (1+\lt\lbt) \, (1 - \beta^2) \, (1 - c^2_\theta) \Big] 
\eeq
where $c_\theta \equiv \cos\theta$ and $s_\theta \equiv \sin\theta$. 

In expressing the $|{\cal M}|^2_{NP}$ for the respective cases, it is 
useful to introduce the notation
\beq
\barr{rclcl c rclcl}
T \equiv 1 - \beta \, c_\theta & = & \dis \frac{2}{s} \, (m_t^2 - t)
& \qquad \quad & 
U \equiv 1 + \beta \, c_\theta & = & \dis \frac{2}{s} \, (m_t^2 - u)
\\[2ex]
\earr
\eeq   

\noindent
\underline{\bf Triplet Diquark Exchange}

With $y^2 = y_S^2 + y_P^2$, we have
\beq
\barr{rcl}
|{\cal M}|^2_{NP} & = & \dis
\dfrac{s^2}{192 \, (u-\mphi^2)^2}
\Big[   y^4 \, U^2
- 4 \, \yS^2 \, \yP^2 \, \lt \, \lbt \, (\beta + c_\theta)^2 
+ 2\, \yS \, \yP \, y^2 \, (\lt - \lbt) \, U \, (\beta + c_\theta) \Big]
\\[3ex]
& + & \dis
\dfrac{g_s^2 \, s}{72 \, (u-\mphi^2)}\Big[y^2 \, 
( 2 - \beta^2 \, c_\theta^2 ) 
\\[1ex]
& & \dis \hspace*{5em}
- \lt\, \lbt \, y^2 \, (  2 \, c_\theta^2 + \beta^2 \, s_\theta^2 )
+2 \, \yS \, \yP \, (\lt-\lbt) \, (\beta + 2c_\theta + \beta c^2_\theta)   \Big] 
\ .
\earr
\label{diq:mesq}
\eeq
For the sextet diquark, only the color factors would change~\cite{Tait}.

\newpage
\noindent
\underline{\bf $Z'$ Exchange}

Defining $x \equiv m_t^2 / 2 \, m_{Z'}^2$, we have
\beq
\barr{rcl}
|{\cal M}|^2_{NP} & = & \dis
\dfrac{\gX^4 \, s^2}{16 \, (t-\mZ^2)^2}
\Big[
\left\{
U^2 
+ 2 \, x \, (1 - \beta^2) 
+ x^2 \,  T^2
\right\}
\\[1ex]
& & \dis \hspace*{5em}
- \lt\lbt
\left\{
(\beta + c_\theta)^2 
+ 2 \, x \, (1 - \beta^2) \, c^2_\theta 
+ x^2 \, (\beta - c_\theta)^2
\right\}
\\[1ex]
&&\hspace*{5em}
+ (\lt - \lbt)
\left\{
(\beta + c_\theta) U
+ 2 \, x \, (1 - \beta^2)  \, c_\theta
- x^2 \, (\beta - c_\theta) T
\right\}
\Big]
\\[2ex]
&-&
 \dfrac{g_s^2\gX^2 \, s}{18 \,(t-\mZ^2)}
\Big[
\left\{U^2 + 1 - \beta^2
+ x \, \left( T^2 + 1 - \beta^2 \right)
\right\}
 \\
& & \dis \hspace*{5em}
- \lt\lbt
\left\{ (1 + x) \, (2 \, c_\theta^2 + \beta^2 \, s_\theta^2) 
+ 2 \, (1 - x) \, \beta \, c_\theta
\right\}
 \\
&& \dis \hspace*{5em}
+ (\lt - \lbt)
\left\{
\beta \, (1 - x) \, (1 + c^2_\theta) + 2 \, (1 + x) \, c_\theta
\right\}
\Big] \ .
\earr
\label{zpr:mesq}
\eeq

\noindent
\underline{\bf Flavor Non-Universal Axigluon Exchange}

Defining combinations of couplings as 
\beq
\barr{rclcrclcrcl}
\sq & \equiv & (\gVq)^2 + (\gAq)^2
& \qquad \quad & 
\pq & \equiv & \gVq \, \gAq
& \qquad \quad & 
\pv & \equiv & \gVq \, \gVt

\\[1.5ex]
{\cal S_T^\pm} & \equiv & (\gVt)^2 \pm (\gAt)^2
& \qquad \quad & 
\pt & \equiv & \gVt \, \gAt
& \qquad \quad & 
\pa & \equiv & \gAq \, \gAt
\earr
\eeq
we have
\beq
\barr{rcl}
|{\cal M}|^2_{NP} & = & \dis 
\dfrac{s\, {\cal J}_A}{18 \, [(s-\mA^2)^2 + \gA^2\mA^2]} \ ,
\earr
\eeq
where 
\beq
\barr{rcl}
{\cal J}_A & = & \dis
s \, 
\Big[
\sq \, \stp \, (1 + \beta^2 c^2_\theta) 
+ \sq \, \stm \, (1 - \beta^2) 
+ 8 \, \pq \, \pt \, \beta \, c_\theta 
\\[2ex]
&& \dis \hspace*{2em} - \lt\lbt \,
\Big\{
\sq \, \stp \, (\beta^2 + c^2_\theta) 
+ \sq \, \stm \, (1 - \beta^2) \, c^2_\theta 
+ 8 \, \pq \, \pt \, \beta \, c_\theta  
\Big\}
\\[2ex]
&&\dis \hspace*{2em} 
+ (\lt-\lbt)
\Big\{
2 \, \sq \, \pt \, \beta \, (1 + c^2_\theta) 
+ 4 \, \pq \, \Big((\gVt)^2 + \beta^2(\gAt)^2\Big) \, c_\theta 
\Big\}
\Big]
\\[3ex]
&+&
g_s^2 \, (s-\mA^2) \, 
\Big[
\left\{
4 \, \pa \, \beta \, c_\theta 
+ 2 \pv \, (2 - \beta^2 \, s^2_\theta )\right\}  \\
&&\dis \hspace*{8em}
- \lt\lbt
\Big\{
4 \, \pa \, \beta \, c_\theta + 
2 \, \pv \, (2 \, c_\theta^2 + \beta^2 \, s_\theta^2)
\Big\}  \\
&& \dis \hspace*{8em}
+ (\lt-\lbt)
\Big\{4\gAq\gVt c_\theta + 2\gVq\gAt\beta(1 + c^2_\theta)\Big\}
\Big] \ .
\earr
\label{axi:mesq}
\eeq

For brevity's sake, we are not presenting the 
SM expressions for $g g \to t \bar t$.

\newpage



\begin{thebibliography}{99}
\label{sec:references}

\bibitem{Nakamura:2010zzi}
  K.~Nakamura {\it et al.}  [Particle Data Group],
  J.\ Phys.\ G{\bf 37}, 075021 (2010).

\bibitem{CDF_AFB_2008_PRL}
  T.~Aaltonen {\it et al.}  [CDF Collaboration],
  Phys.\ Rev.\ Lett.\  {\bf 101}, 202001 (2008)
  [arXiv:0806.2472 [hep-ex]].

\bibitem{D0_AFB_2008_PRL}
  V.~M.~Abazov {\it et al.}  [D0 Collaboration],
  Phys.\ Rev.\ Lett.\  {\bf 100}, 142002 (2008)
  [arXiv:0712.0851 [hep-ex]].

\bibitem{CDF_AFB_2009}
CDF Conference Note 9724, \\
http://www-cdf.fnal.gov/physics/new/top/public\_tprop.html

\bibitem{CDF_AFB_2010}
CDF Conference Note 10185, \\
http://www-cdf.fnal.gov/physics/new/top/public\_tprop.html

\bibitem{D0_AFB_2010}
D0 Conference Note 6062,\\
http://www-d0.fnal.gov/Run2Physics/WWW/results/prelim/TOP/T90/

\bibitem{Abazov:2010hv}
  V.~M.~Abazov {\it et al.}  [D0 Collaboration],
  Phys.\ Rev.\  D{\bf 82} (2010) 032001
  [arXiv:1005.2757 [hep-ex]].

\bibitem{top_reviews}
For a review, see for example,\\ 
  W.~Bernreuther,
  J.\ Phys.\ G{\bf 35}, 083001 (2008)
  [arXiv:0805.1333 [hep-ph]];
\\
  T.~Han,
  Int.\ J.\ Mod.\ Phys.\  A{\bf 23}, 4107 (2008)
  [arXiv:0804.3178 [hep-ph]].

\bibitem{AFB_SMNLO}
  J.~H.~Kuhn and G.~Rodrigo,
  Phys.\ Rev.\ Lett.\  {\bf 81}, 49 (1998)
  [arXiv:hep-ph/9802268];
  Phys.\ Rev.\  D {\bf 59}, 054017 (1999)
  [arXiv:hep-ph/9807420].

\bibitem{AFB_SM_others}
  M.~T.~Bowen, S.~D.~Ellis and D.~Rainwater,
  Phys.\ Rev.\  D{\bf 73}, 014008 (2006)
  [arXiv:hep-ph/0509267];
\\
  L.~G.~Almeida, G.~F.~Sterman and W.~Vogelsang,
  Phys.\ Rev.\  D{\bf 78}, 014008 (2008)
  [arXiv:0805.1885 [hep-ph]].

\bibitem{Sehgal:1987wi}
 L.~M.~Sehgal and M.~Wanninger,
 Phys.\ Lett.\  B{\bf 200}, 211 (1988).

\bibitem{DC&RMG}
  D.~Choudhury, R.~M.~Godbole, R.~K.~Singh and K.~Wagh,
  Phys.\ Lett.\  B {\bf 657}, 69 (2007)
  [arXiv:0705.1499 [hep-ph]].

\bibitem{Murayama}
  S.~Jung, H.~Murayama, A.~Pierce and J.~D.~Wells,
  Phys.\ Rev.\  D {\bf 81}, 015004 (2010)
  [arXiv:0907.4112 [hep-ph]].

\bibitem{Tait}
  J.~Shu, T.~M.~P.~Tait and K.~Wang,
  Phys.\ Rev.\  D {\bf 81}, 034012 (2010)
  [arXiv:0911.3237 [hep-ph]].

\bibitem{Frampton}
  P.~H.~Frampton, J.~Shu and K.~Wang,
  Phys.\ Lett.\  B {\bf 683}, 294 (2010)
  [arXiv:0911.2955 [hep-ph]].

\bibitem{AFB_others}
  C.~H.~Chen, G.~Cvetic and C.~S.~Kim,
  arXiv:1009.4165 [hep-ph];
\\
  Y.~K.~Wang, B.~Xiao and S.~H.~Zhu,
  arXiv:1008.2685 [hep-ph];
\\
  B.~Xiao, Y.~K.~Wang and S.~H.~Zhu,
  Phys.\ Rev.\  D {\bf 82}, 034026 (2010)
  [arXiv:1006.2510 [hep-ph]];
\\
  Q.~H.~Cao, D.~McKeen, J.~L.~Rosner, G.~Shaughnessy and C.~E.~M.~Wagner,
  Phys.\ Rev.\  D {\bf 81}, 114004 (2010)
  [arXiv:1003.3461 [hep-ph]];
\\
  V.~Barger, W.~Y.~Keung and C.~T.~Yu,
  Phys.\ Rev.\  D {\bf 81}, 113009 (2010)
  [arXiv:1002.1048 [hep-ph]];
\\
  J.~Cao, Z.~Heng, L.~Wu and J.~M.~Yang,
  Phys.\ Rev.\  D {\bf 81}, 014016 (2010)
  [arXiv:0912.1447 [hep-ph]];
\\
  A.~Arhrib, R.~Benbrik and C.~H.~Chen,
  Phys.\ Rev.\  D {\bf 82}, 034034 (2010)
  [arXiv:0911.4875 [hep-ph]];
\\
  K.~Cheung, W.~Y.~Keung and T.~C.~Yuan,
  Phys.\ Lett.\  B {\bf 682}, 287 (2009)
  [arXiv:0908.2589 [hep-ph]];
\\
  A.~Djouadi, G.~Moreau, F.~Richard and R.~K.~Singh,
  Phys.\ Rev.\  D {\bf 82}, 071702 (2010)
  [arXiv:0906.0604 [hep-ph]];
\\
  C.~Degrande, J.~M.~Gerard, C.~Grojean, F.~Maltoni and G.~Servant,
  arXiv:1010.6304 [hep-ph];
\\
  M.~V.~Martynov and A.~D.~Smirnov,
  arXiv:1010.5649 [hep-ph];
\\
  I.~Dorsner, S.~Fajfer, J.~F.~Kamenik and N.~Kosnik,
  Phys.\ Rev.\  D {\bf 81}, 055009 (2010)
  [arXiv:0912.0972 [hep-ph]];
\\
  G.~Burdman, L.~de Lima and R.~D.~Matheus,
  arXiv:1011.6380 [hep-ph];
\\
  E.~Alvarez, L.~Da Rold and A.~Szynkman,
  arXiv:1011.6557 [hep-ph];
\\
  J.~A.~Aguilar-Saavedra,
  Nucl.\ Phys.\  B {\bf 843}, 638 (2011)
  [arXiv:1008.3562 [hep-ph]].

\bibitem{Chivukula:2010fk}
  R.~S.~Chivukula, E.~H.~Simmons and C.~P.~Yuan,
  Phys.\ Rev.\  D {\bf 82}, 094009 (2010)
  [arXiv:1007.0260 [hep-ph]].

\bibitem{Jung:2009pi}
  D.~W.~Jung, P.~Ko, J.~S.~Lee and S.~h.~Nam,
  Phys.\ Lett.\  B {\bf 691}, 238 (2010)
  [arXiv:0912.1105 [hep-ph]].

\bibitem{Hikasa:1999wy}
K.~I.~Hikasa, J.~M.~Yang, B.~L.~Young,
Phys.\ Rev.\  {\bf D60}, 114041 (1999), [hep-ph/9908231].

\bibitem{toppol_bsm}
  E.~Boos, H.~U.~Martyn, G.~A.~Moortgat-Pick, M.~Sachwitz, A.~Sherstnev and P.~M.~Zerwas,
  Eur.\ Phys.\ J.\  C {\bf 30}, 395 (2003)
  [arXiv:hep-ph/0303110];
\\
  T.~Gajdosik, R.~M.~Godbole and S.~Kraml,
  JHEP {\bf 0409}, 051 (2004)
  [arXiv:hep-ph/0405167];
\\
  M.~Perelstein and A.~Weiler,
  JHEP {\bf 0903}, 141 (2009)
  [arXiv:0811.1024 [hep-ph]];
\\
  M.~M.~Nojiri and M.~Takeuchi,
  JHEP {\bf 0810}, 025 (2008)
  [arXiv:0802.4142 [hep-ph]];
\\
  M.~Arai, K.~Huitu, S.~K.~Rai and K.~Rao,
  JHEP {\bf 1008}, 082 (2010)
  [arXiv:1003.4708 [hep-ph]];
\\
  K.~Huitu, S.~K.~Rai, K.~Rao, S.~D.~Rindani and P.~Sharma,
  arXiv:1012.0527 [hep-ph].

\bibitem{Agashe:2006hk}
  K.~Agashe, A.~Belyaev, T.~Krupovnickas, G.~Perez and J.~Virzi,
  Phys.\ Rev.\  D {\bf 77}, 015003 (2008)
  [arXiv:hep-ph/0612015].

\bibitem{jhepus2}
  P.~S.~Bhupal Dev, A.~Djouadi, R.~M.~Godbole, M.~M.~Muhlleitner and S.~D.~Rindani,
  Phys.\ Rev.\ Lett.\  {\bf 100}, 051801 (2008)
  [arXiv:0707.2878 [hep-ph]].

\bibitem{Godbole:2010kr}
  R.~M.~Godbole, K.~Rao, S.~D.~Rindani and R.~K.~Singh,
  JHEP {\bf 1011}, 144 (2010)
  [arXiv:1010.1458 [hep-ph]];
  AIP Conf.\ Proc.\  {\bf 1200}, 682 (2010)
  [arXiv:0911.3622 [hep-ph]];
  B.~C.~Allanach {\it et al.},
  arXiv:hep-ph/0602198.

\bibitem{Godbole:2006tq}
  R.~M.~Godbole, S.~D.~Rindani and R.~K.~Singh,
  JHEP {\bf 0612}, 021 (2006)
  [arXiv:hep-ph/0605100].

\bibitem{Shelton}
  J.~Shelton,
  Phys.\ Rev.\  D {\bf 79}, 014032 (2009)
  [arXiv:0811.0569 [hep-ph]];
\\
  D.~Krohn, J.~Shelton, L.~T.~Wang,
  JHEP {\bf 1007 } (2010)  041.
  [arXiv:0909.3855 [hep-ph]].

\bibitem{LEP_QCD}
  M.~Jezabek and J.~H.~Kuhn,
  Nucl.\ Phys.\  B {\bf 320}, 20 (1989);
\\
  A.~Czarnecki, M.~Jezabek and J.~H.~Kuhn,
  Nucl.\ Phys.\  B {\bf 351}, 70 (1991);
\\
  A.~Brandenburg, Z.~G.~Si and P.~Uwer,
  Phys.\ Lett.\  B {\bf 539}, 235 (2002)
  [arXiv:hep-ph/0205023].

\bibitem{Hioki_et_al}
  B.~Grzadkowski and Z.~Hioki,
  Phys.\ Lett.\  B {\bf 476}, 87 (2000)
  [arXiv:hep-ph/9911505];
  Phys.\ Lett.\  B {\bf 557}, 55 (2003)
  [arXiv:hep-ph/0208079];
  Phys.\ Lett.\  B {\bf 529}, 82 (2002)
  [arXiv:hep-ph/0112361];
\\
  Z.~Hioki,
  arXiv:hep-ph/0210224;
  arXiv:hep-ph/0104105;
\\
  K.~Ohkuma,
  Nucl.\ Phys.\ Proc.\ Suppl.\  {\bf 111}, 285 (2002)
  [arXiv:hep-ph/0202126].

\bibitem{Rindani}
  S.~D.~Rindani,
  Pramana {\bf 54}, 791 (2000)
  [arXiv:hep-ph/0002006].

\bibitem{jhepus1}
  R.~M.~Godbole, S.~D.~Rindani and R.~K.~Singh,
  Phys.\ Rev.\  D {\bf 67}, 095009 (2003)
  [Erratum-ibid.\  D {\bf 71}, 039902 (2005)]
  [arXiv:hep-ph/0211136].

\bibitem{LHC_spin_plans}
CERN Notes, http://cdsweb.cern.ch/record/814352; http://cdsweb.cern.ch/record/973111.

\bibitem{CDF_spincorr}
CDF Conference Note 10211, \\
http://www-cdf.fnal.gov/physics/new/top/confNotes/

\bibitem{D0_spincorr}
D0 Conference Note 5950, \\
http://www-d0.fnal.gov/Run2Physics/WWW/results/prelim/TOP/T84/

\bibitem{Hewett:1988xc}
  J.~L.~Hewett and T.~G.~Rizzo,
  Phys.\ Rept.\  {\bf 183}, 193 (1989).

\bibitem{Frampton&Glashow}
  P.~H.~Frampton and S.~L.~Glashow,
  Phys.\ Lett.\  B {\bf 190}, 157 (1987);
  Phys.\ Rev.\ Lett.\  {\bf 58}, 2168 (1987).

\bibitem{CDF_dijet_2008}
  T.~Aaltonen {\it et al.}  [CDF Collaboration],
  Phys.\ Rev.\  D {\bf 79}, 112002 (2009)
  [arXiv:0812.4036 [hep-ex]].

\bibitem{Antunano:2007da}
  O.~Antunano, J.~H.~Kuhn and G.~Rodrigo,
  Phys.\ Rev.\  D {\bf 77}, 014003 (2008)
  [arXiv:0709.1652 [hep-ph]].

\bibitem{CDF_NP_in_ttbar}
  T.~Aaltonen {\it et al.}  [CDF Collaboration],
  Phys.\ Lett.\  B {\bf 691}, 183 (2010)
  [arXiv:0911.3112 [hep-ex]].

\bibitem{Choudhury:2005dg}
  D.~Choudhury, M.~Datta and M.~Maity,
  Phys.\ Rev.\  D {\bf 73}, 055013 (2006)
  [arXiv:hep-ph/0508009].

\bibitem{CDF_mtt}
  T.~Aaltonen {\it et al.}  [CDF Collaboration],
  Phys.\ Rev.\ Lett.\  {\bf 102}, 222003 (2009)
  [arXiv:0903.2850 [hep-ex]].

\bibitem{CTEQ}
  J.~Pumplin, D.~R.~Stump, J.~Huston, H.~L.~Lai, P.~M.~Nadolsky and W.~K.~Tung,
  JHEP {\bf 0207}, 012 (2002)
  [arXiv:hep-ph/0201195].

\bibitem{CDF_csec}
CDF Conference Note 9913, \\
http://www-cdf.fnal.gov/physics/new/top/public\_xsection.html

\bibitem{K-factor}
  M.~Cacciari, S.~Frixione, M.~L.~Mangano, P.~Nason and G.~Ridolfi,
  JHEP {\bf 0809}, 127 (2008)
  [arXiv:0804.2800 [hep-ph]];\\
\textit{See also,}\\
  S.~Moch and P.~Uwer,
  Phys.\ Rev.\  D {\bf 78}, 034003 (2008)
  [arXiv:0804.1476 [hep-ph]]; \\
  N.~Kidonakis and R.~Vogt,
  Phys.\ Rev.\  D {\bf 78}, 074005 (2008)
  [arXiv:0805.3844 [hep-ph]].

\bibitem{CMS_ttbar}
  V.~Khachatryan {\it et al.}  [CMS Collaboration],
  arXiv:1010.5994 [hep-ex].

\bibitem{ATLAS_ttbar}
  ATLAS Collaboration,
  arXiv:1012.1792 [hep-ex].

\bibitem{Bhattacharyya:1995bw}
   G.~Bhattacharyya, D.~Choudhury and K.~Sridhar,
   Phys.\ Lett.\  B {\bf 355}, 193 (1995)
   [arXiv:hep-ph/9504314].

\bibitem{Cao:2010nw}
  J.~Cao, L.~Wu and J.~M.~Yang,
  arXiv:1011.5564 [hep-ph].

\bibitem{Jung:2010yn}
  D.~W.~Jung, P.~Ko and J.~S.~Lee,
  arXiv:1011.5976 [hep-ph].

\bibitem{Aaltonen:2011kc}
  T.~Aaltonen {\it et al.}  [The CDF Collaboration],
  arXiv:1101.0034 [hep-ex].

\end{thebibliography}
\end{document}